\documentclass[11pt,a4paper]{article}
\pdfoutput=1

\usepackage{jheppub}
\usepackage{latexsym}
 \usepackage{multirow}
\usepackage{color}
\usepackage[usenames,dvipsnames,table]{xcolor}
\usepackage{graphicx}
\usepackage{epsfig}  
 \usepackage{epsf}    
\usepackage{dcolumn}
\usepackage{bm}
\usepackage{dcolumn}
\usepackage{textcomp}
\usepackage{float}
\usepackage{subfig}
\usepackage{hypcap}
\usepackage[normalem]{ulem}
\usepackage{amsmath}
\newcommand{\stkout}[1]{\ifmmode\text{\sout{\ensuremath{#1}}}\else\sout{#1}\fi}
\usepackage[]{hyperref}
\usepackage{color}
\usepackage{pifont}

\usepackage{url}
\hypersetup{
  bookmarks=true,         
  unicode=false,          
  pdftoolbar=true,        
 pdfmenubar=true,        
 pdffitwindow=true,     
 pdfstartview={FitH},    
 pdfsubject={Two Higgs doublet},   
 pdfnewwindow=true,      
 pdfcreator={RevTeX},
 colorlinks=true,       
 linkcolor=red,          
 citecolor=blue,        
 filecolor=black,      
 urlcolor=blue,           
  }

\newcommand{\be}{\begin{equation}}
\newcommand{\ee}{\end{equation}}
\newcommand{\ba}{\begin{eqnarray}}
\newcommand{\ea}{\end{eqnarray}}

\newcommand{\D}{\Delta}

\makeatletter
\renewcommand{\fnum@table}{\textbf{\tablename~\thetable}}
\renewcommand{\fnum@figure}{\textbf{\figurename~\thefigure}}
\makeatother

\title{Minimal and non-minimal Universal Extra Dimension models in the light of LHC data at 13 TeV}
\author[a,b]{Avnish}
\author[a,b]{Kirtiman Ghosh}
\author[a,b]{Tapoja Jha}
\author[c]{Saurabh Niyogi}
\affiliation[a]{Institute of Physics, Sachivalaya Marg, Bhubaneswar, Odisha 751005, India}
\affiliation[b]{Homi Bhabha National Institute, Training School Complex, Anushakti Nagar, Mumbai 400085, India}
\affiliation[c]{Gokhale Memorial Girls' College, 1/1 Harish Mukherjee Road, Kolkata 700 020, India}

\emailAdd{avnish@iopb.res.in}
\emailAdd{kirti.gh@gmail.com}
\emailAdd{tapoja.phy@gmail.com}
\emailAdd{~~~~~~~saurabhphys@gmail.com}

\abstract
    {Universal Extra Dimension (UED) is a well-motivated and well-studied scenario. One of the main motivations is the presence of a dark matter (DM) candidate namely, the lightest level-1 Kaluza-Klein (KK) particle (LKP), in the particle spectrum of UED. The minimal version of UED ({\em mUED}) scenario is highly predictive with only two parameters namely, the radius of compactification and cut-off scale, to determine the phenomenology. Therefore, stringent constraint results from the WMAP/PLANCK measurement of DM relic density (RD) of the universe. The production and decays of level-1 quarks and gluons in UED scenarios give rise to multijet final states at the Large Hadron Collider (LHC) experiment. We study the ATLAS search for multijet plus missing transverse energy signatures at the LHC with 13 TeV center of mass energy and 139 inverse femtobarn integrated luminosity. In view of the fact that the DM RD allowed part of {\em mUED} parameter-space has already been ruled out by the ATLAS multijet search, we move on to a less restricted version of UED namely, the non-minimal UED ({\em nmUED}), with non-vanishing boundary-localized terms (BLTs). The presence of BLTs significantly alters the dark matter as well as the collider phenomenology of {\em nmUED}. We obtain stringent bounds on the BLT parameters from the ATLAS multijet plus missing transverse energy search.}

\preprint{IP/BBSR/2020-09}
    

\begin{document}
\maketitle
\flushbottom

\section{Introduction}
\label{intro}

After almost a decade long running, the Large Hadron Collider (LHC) collected and analyzed 139 fb$^{-1}$ integrated luminosity data along with a boasting discovery of the Higgs Boson \cite{Aad:2012tfa,Chatrchyan:2012xdj} confirming the mechanism behind masses of the weak gauge bosons and fermions of the Standard Model (SM). Numerous analysis of the LHC data in a variety of channels establish the predictions of the SM on firm footing \cite{Chatrchyan:2012xdj}. Nonetheless, the existence of Higgs boson brings forth other questions in terms of the stability of its mass {\em etc.} On the experimental front, the evidence of neutrino oscillation and hence, the presence of tiny neutrino masses casts shadow over the SM. {Although one can solve the problem by incorporating right-handed gauge singlet neutrinos and assigning additional Yukawa terms in the SM, provided the mass hierarchy in the SM fermion sector is acceptable.} A rather more daunting task is to incorporate the
idea of a new weakly interacting massive particle in the theory in order to explain certain pressing cosmological as well 
as astronomical evidences in the name of dark matter (DM). Such inadequacies of the SM lead to 
plenty of novel ideas which are either extension of the SM or, a completely new theory which would come down to the SM at an appropriate limit.

The invocation of theories with extra spatial dimension(s) are of interest for a number of reasons. The most profound ones are the
stability of the Higgs boson mass and the related hierarchy problems which were successfully explained by Arkani-Hamed, Dimopoulos, Dvali (ADD) model
\cite{ArkaniHamed:1998rs,ArkaniHamed:1998nn} and later by Randall-Sundrum (RS) \cite{Randall:1999ee,Randall:1999vf} model. Extra dimensional theories can also achieve a light neutrino without introducing any heavy mass
scale \cite{Dienes:1998sb}, the unification of gauge couplings \cite{Dienes:1998vh} and can also account for hierarchies present in the SM fermion masses \cite{ArkaniHamed:1999dc}. Among a variety of extra dimensional frameworks, we confine ourselves to {a particular variant}, {called} the {\em Universal Extra Dimension (UED)} model(s) where all the SM fields are allowed to propagate into the space(s) beyond the usual 3+1 dimensional space-time\cite{ANTONIADIS1990377,PhysRevD.64.035002,PhysRevD.66.056006}.
Of course, there are other prospects of working with such frameworks such as, electroweak symmetry breaking without invoking
a fundamental scalar \cite{ArkaniHamed:2000hv}, a cosmologically viable dark matter candidate \cite{Servant:2002aq,Kakizaki:2006dz}, {a unification scale at a few TeV \cite{Dienes:1998vg}, an explanation for the long life-time of proton \cite{Appelquist:2001mj} and the number of fermion
generations to be an integral multiple of three \cite{Dobrescu:2001ae}}, and above all, 
a chance to probe the model at collider experiments with its promising signatures \cite{Appelquist:2000nn,Belyaev:2012ai,Kakuda:2013kba,Belanger:2012mc,Ghosh:2018mck,Ghosh:2008dp,Ghosh:2008ix,Beuria:2017jez,Dey:2014ana,Edelhauser:2013lia,Flacke:2012ke,Huang:2012kz,Flacke:2011nb,Nishiwaki:2011gm,Murayama:2011hj,Choudhury:2011jk,Ghosh:2010tp,Bhattacherjee:2010vm,Bertone:2010ww,Freitas:2007rh,Dobrescu:2007ec,Macesanu:2002db,Ghosh:2012zc,Choudhury:2009kz,Rizzo:2001sd,Muck:2003kx,Bhattacharyya:2005vm,Battaglia:2005zf,Bhattacherjee:2005qe,Datta:2005zs,Datta:2005vx}.

{In this work,} we {study the collider phenomenology of} {a couple of} simple variants of UED {scenarios} {which are characterized by a single flat universal (accessible to all the SM particles) extra dimension ($y$), compactified on a $S_1/Z_2$ orbifold with radius $R$ {({\em oneUED} scenarios)}.} {In particular, we consider both the {\em minimal (mUED)} and {\em non-minimal (nmUED)} versions of the {\em oneUED} model. The particle spectrum of {\em oneUED} contains infinite towers of Kaluza-Klein (KK) modes (identified by an integer $n$, called the KK-number) for each of the SM fields. The zero modes are identified as the corresponding SM particles. From a 4-dimensional perspective, the conservation of the momentum along fifth direction  implies conservation of the KK-number. However, the additional $Z_2$ symmetry $(y \leftrightarrow -y$), which is required to obtain chiral structure of the SM fermions, breaks the translational invariance along the 5$^{\rm th}$ dimension. As a result, KK-number conservation breaks down at loop-level, leaving behind only a conserved KK-parity, defined as $(-1)^n$, which is an automatic outcome of the $S_1/Z_2$ orbifolding and has several interesting consequences. For example, KK-parity ensures the stability of the lightest KK particle (LKP), allows only the pair productions of level-1 KK particles at the collider, and prohibits KK modes from affecting electroweak (EW) precision observables at tree-level. }

{{\em OneUED}, being a higher dimensional theory, is non-renormalizable and hence, should be treated as an effective theory valid upto a cut-off scale $\Lambda$. Apart from the usual the SM kinetic, Yukawa and scalar potential terms for the 5D fields, the {\em oneUED} Lagrangian also includes additional SM gauge and Lorentz invariant terms like, the vector-like bulk mass terms \cite{Kong:2010qd,Flacke:2011nb,Chen:2009gz,Huang:2012kz,Park:2009cs} for the 5D fermions. Furthermore, one can, in principle, also add kinetic (and Yukawa) terms\footnote{{These terms are known as boundary localized terms (BLTs). It is important to note that the BLTs are only consistent with 4D Lorentz symmetry as well as the gauge symmetry.}} for all the 5D fields at the orbifold fixed points, {\em i.e.,} the boundaries of the bulk and the brane \cite{delAguila:2003bh,Carena:2002me}. The parameters associated with the BLTs are not a priory known quantities (since they are related to ultra-violet (UV) completion for such scenarios) and thus, would serve as extra free parameters of the theory. In the {\em minimal} version of {\em oneUED} \cite{Cheng:2002iz}, all BLTs are assumed to vanish at the cut-off scale ($\Lambda$)
  and are radiatively generated at the low scale which ultimately  appear as corrections to the masses of the KK particles.}
Therefore, {in addition to the SM parameters, the phenomenology of {\em mUED}} is determined by only two additional parameters namely, the radius of compactification, $R$ and the cut-off scale, $\Lambda$. Hence, it's predictions are very specific and easily testable at different high energy physics (HEP) experiments. {As a result,} verdicts from {different non-collider 
and collider} experiments, {like the LHC and various DM experiments} can {easily} rule out {{\em mUED}. It has already been shown in the literature \cite{Choudhury:2016tff,Deutschmann:2017bth,Cornell:2014jza} that the parts of $R^{-1}$--$\Lambda$ plane of {\em mUED} which are consistent with the WMAP/PLANCK \cite{Komatsu:2010fb,Ade:2015xua} observed relic density (RD) data, are on the verge of being excluded from the direct searches for the KK particles at the LHC.} {This motivates us to} move on to a less stricter version of {\em oneUED} with more parameters {namely, the BLT parameters}. This is where the {\em non-minimal UED} comes into the picture. In {\em nmUED}, BLT parameters give rise to  modifications in
{the KK particle} masses as well as {interactions} \cite{Flacke:2008ne,Datta:2012tv,Flacke:2013nta,Datta:2013yaa,Flacke:2013pla,Shaw:2017whr}. The effect of such alterations is rather dramatic at the colliders as well as at the dark matter experiments.
Studies on various phenomenological aspects of {\em nmUED} including the LHC phenomenology are abundant in number. In particular, results from the Higgs boson data \cite{Datta:2013ufa,Dey:2013cqa,Ghosh:2014uwa} and different DM experiments \cite{Datta:2013nua,Flacke:2017xsv} directly constrain the model. Allowed parameter space in accordance with constraints of flavor physics are also obtained \cite{Datta:2015aka,Datta:2016flx,Biswas:2017vhc}. Apart from these, a number of theoretical constraints are also placed from unitarity  bounds \cite{Jha:2016sre}, Z $\rightarrow$ bb decay width \cite{Jha:2014faa, Jha:2017oek}, flavor physics \cite{Dasgupta:2018nzt}, collider phenomenology \cite{Ganguly:2018pzs}, and others \cite{Dey:2016cve,Chiang:2018oyd,Datta:2014yua} as well.

{In this article, we have studied the collider signatures of {\em mUED} and {\em nmUED} in the context of recent LHC searches for beyond the SM (BSM) scenarios. The level-1 KK particles are expected to be in the mass range of few hundreds of GeV to few TeVs. Being strongly interacting, the level-1 KK quarks (both the singlet, $q^{(1)}$, and doublet,
$Q^{(1)}$), and gluons, $G^{(1)}$, can be copiously produced in pairs at the LHC. These, subsequently, decay into the SM particles and the LKP via cascades involving other level-1 KK particles. Therefore, the pair productions of the level-1 KK particles give rise to generic multijet $+$ multilepton $+$ missing transverse energy\footnote{{The LKP, being stable and weakly interacting, escapes the LHC detectors and thus, contributes to the missing energy signature.}} ($E_T\!\!\!\!\!\!/~$) signatures at the LHC.} 

Now turning on to the actual ambit of our work, it is worth mentioning that the LHC collaborations have so far performed
dedicated analysis in the multijet as well as multilepton channels {in the context of supersymmetric and other BSM scenarios. In particular, the ATLAS collaboration have studied the signatures 
of gluino and/or squark (supersymmetric partner of gluon and quark, respectively) pair productions in multijet plus missing transverse energy channels at the LHC at 13 TeV center of mass energy with 139 fb$^{-1}$ integrated luminosity.}   
Non-observation of expected signal (over background) results in strong constraints on many sparticle (supersymmetric particle) masses. One can always perform the ditto analysis as done by the ATLAS for any model to constrain the parameter space of that particular model from the LHC data. In this article, we follow this well trodden path and revisit the status of the {\em mUED} and {\em nmUED} scenarios after the LHC run-II data.

The paper is organized as follows. In the following section \ref{Collider_Phenomenology}, we describe the ATLAS multijet analysis strategy and 
 validate our methodology by reproducing the ATLAS results. Next, we first look for the status of the minimal version of Universal Extra Dimension under the lens of LHC data collected at 13 TeV.
In Section \ref{nmUED_model}, we describe the non-minimal UED model. Section \ref{nmued_collider} comprises of the LHC phenomenology of the {\em nmUED} model followed by the concluding remarks in section \ref{conclusion}.

 \section{Collider Phenomenology}\label{Collider_Phenomenology}
 We {have} closely followed the latest ATLAS $n_j +  E_T\!\!\!\!\!\!/~$ \cite{ATLAS-CONF-2019-040} search with 139~fb$^{-1}$ integrated luminosity data at the 13 TeV LHC. {Although, the analysis in Ref.~\cite{ATLAS-CONF-2019-040} is dedicated for the search of squarks and gluinos in the context of supersymmetry, the model independent 95\% CL upper limits on the visible $n_j +E_T\!\!\!\!\!\!/~$ cross-sections ($\langle \epsilon \sigma\rangle_{obs}^{95}$) for 
 different signal regions (SRs), can be used to constrain the parameter space of other BSM scenarios which also give rise to similar final states. A brief description about the reconstruction of various objects (jets, leptons {\em etc.}), event selection criteria,
 definition of different SRs are presented in the following.}\\
\noindent{\bf \em Object Reconstruction:} Jet candidates have been reconstructed using anti-$k_T$ \cite{Cacciari:2008gp} algorithm implemented in FastJet \cite{Cacciari:2011ma} with jet radius parameter $0.4$. Reconstructed jets with $p^j_T >  20$ GeV and $|\eta^j| < 2.8$ are considered for further analysis. Electron (muon) candidates are required to have $p^l_T > 7 (6)$ GeV and within $|\eta^l| < 2.47 (2.7)$.  Next, the overlapping between identified leptons and jets in the final state are resolved by discarding any electron/muon candidate lying within a distance $\Delta \rm{R} < \rm{min}(0.4,0.04+\frac{10 \rm{ GeV}}{p_T^{e/\mu}})$ of any reconstructed jet  candidate. Missing transverse momentum vector $p_T^{\rm mis}$ (with magnitude $E_T\!\!\!\!\!\!/~$) is reconstructed using all remaining visible entities, viz. jets, leptons, photons and all calorimeter clusters not associated to such objects. For a signal having $n_j$ jets, effective mass $\rm{m}_{eff}$ is defined as the scalar sum of $E_T\!\!\!\!\!\!/~$ and
  the transverse momenta of all the $n_j$ jet candidates having $p_T > 50$ GeV. Whereas, $\rm{H}_T $ is calculated as the scalar sum of transverse momentum of all jets with $p_T>50$ GeV and $|\eta|<2.8$. After reconstructing different physics objects, events are pre-selected for further analysis. As SUSY and other BSM scenarios are expected to reside in the high mass scale region, events are preselected accordingly and thus, in the process, unnecessary events are rejected. The preselection criteria is summarized below.\\
\noindent {\bf \em Preselection criteria:} {Events containing a leading jet with $p^{j_1}_T > 200$ GeV and atleast a second jet with  $p^{j_2}_T > 50$ GeV are considered for further analysis. Only zero lepton events are considered {\em i.e.,} events with an isolated  electron (muon) with $p_T > 7 (6)$ GeV are vetoed. 
Events are required to have sufficiently large missing transverse energy ($E_T\!\!\!\!\!\!/~ > 300$ GeV) and effective mass ($ \rm{m}_{\text{eff}} > 800$ GeV) in order to be considered for further analysis. Events failing to satisfy $\Delta \phi(j_{1,2},p^{\text{mis}}_T)_{\text{min}} > 0.4$\footnote{{$\Delta \phi(j_{i},p^{\text{mis}}_T)_{\text{min}}$ is the azimuthal angle between the $i^{\rm th}$ jet and missing transverse momentum vector $p^{\text{mis}}_T$. Jets are ordered according to their $p_T$ hardness ($p^{j_1}_T~>~p^{j_2}_T~>~.....$).}} are also rejected.}\\
\noindent {\bf \em Event Selection and Signal Regions (SRs):} {To make the search process exhaustive, the ATLAS collaboration {\cite{ATLAS-CONF-2019-040}} have defined various signal regions (SRs). Each signal region is designed to study a particular region of parameter space and hence, the signal regions are made mutually exclusive as far as possible. Number of jets sets a powerful criterion in achieving this.} For instance, {in the context of supersymmetry,} a pair of gluinos typically
give more number of jets than squarks in their usual decay modes. Thus, binning different numbers for jets is the first step for segmenting SRs. Moreover, mass splitting of the parent and daughter determines the kinematics of the events. Thus, in addition, specific {cuts on different} kinematic variables (dubbed as multi-bin search) {have also been} applied {to} target specific {mass hierarchies among different BSM particles}. {In Ref.~\cite{ATLAS-CONF-2019-040}, ATLAS collaboration have defined ten signal regions for their model independent study of multijet plus missing energy signatures at the LHC running at 13 TeV center of mass energy with 139 fb$^{-1}$ integrated luminosity. The signal regions are defined by varying numbers of jet multiplicities (between 2--6) along with the minimum value of the effective mass $\rm{m}_{\text{eff}}$. In view of the high level of agreement between predicted background and observed yield in all signal regions, a model independent 95\% CL upper limit is set on the visible BSM contribution to the multijet cross-section ($\langle \epsilon \sigma\rangle_{obs}^{95}$) for each signal region. In our analysis, we have used the ATLAS derived bounds on $\langle \epsilon \sigma\rangle_{obs}^{95}$ in each signal region to constrain the parameter space of {\em mUED} and {\em nmUED}. For the sake of completeness, we have summarized the definitions of few relevant signal regions in Table~\ref{SRs}\footnote{For more details, we refer the reader to Table~8, 9 and 12 of Ref.~\cite{ATLAS-CONF-2019-040}.}.}
\begin{table}[ht!]
  \centering
  \begin{tabular}{|c|c|c|c|c|c|}
    \hline\hline
    &SR2j-1.6  &SR2j-2.2  &SR2j-2.8 &SR4j-1.0 &SR4j-2.2 \\
    \hline\hline
    $\rm{n}_j$ & $\geq 2$ & $\geq 2$ & $\geq 2$ & $\geq 4$ & $\geq 4$ \\
    $p_T(j_1)$[GeV] &$>250$ &$>600$ &$>250$ &$>200$ &$>200$ \\
    $p_T(j_{i=2,..,n_{\rm min}})$[GeV] &$>250$ &$>50$ &$>250$ &$>100$ &$>100$ \\
    $|\eta(j_{1,..,n_{min}})|$ &$<2.0$ &$<2.8$ &$<1.2$ &$<2.0$ &$<2.0$ \\
    $\Delta\phi(j_{1,2,(3)},p_T^{\rm mis})_{\rm min}$ &$>0.8$ &$>0.4$ &$>0.8$ &$>0.4$ &$>0.4$ \\
    $\Delta\phi(j_{i>3},p_T^{\rm mis})_{\rm min}$ &$>0.4$ &$>0.2$ &$>0.4$ &$>0.4$ &$>0.4$\\
    $\rm{Aplanarity}$ &- &- &- &$>0.04$ &$>0.04$\\
    $\frac{E_T\!\!\!\!\!\!/~}{\sqrt{H_T}} [\sqrt{\rm{GeV}}]$ &$>16$ &$>16$ &$>16$ &$>16$ &$>16$ \\
    $\rm{m}_{\rm{eff}}$[TeV] &$>1.6$ &$>2.2$ &$>2.8$ &$>1.0$ &$>2.2$\\
    \hline\hline
    $\langle \epsilon \sigma\rangle_{obs}^{95}$ [fb] & 1.46 & 0.78 & 0.13 & 0.54 & 0.14 \\\hline\hline
    \multicolumn{6}{|c|}{{\em mUED} predictions [fb]}\\\hline\hline
    BP$_1^{m}$ & 0.35 & \cellcolor [HTML]{DFDDDD} 0.92 & 0.03 & 0.28 & 0.03\\
    BP$_2^{m}$ & 0.96 & 0.29 & 0.05 & \cellcolor [HTML]{DFDDDD} 0.58 & 0.11 \\
    BP$_3^{m}$ & \cellcolor [HTML]{DFDDDD} 5.25 & \cellcolor [HTML]{DFDDDD} 2.96 & \cellcolor [HTML]{DFDDDD} 0.33 & \cellcolor [HTML]{DFDDDD} 6.41 & \cellcolor [HTML]{DFDDDD} 0.75 \\\hline\hline
    \multicolumn{6}{|c|}{{\em nmUED} predictions [fb]}\\\hline\hline
    BP$_1^{nm}$ & \cellcolor [HTML]{DFDDDD}  1.67 &  0.61 &  0.08 &  0.05 & 0.03 \\
    BP$_2^{nm}$ & 0.45 &\cellcolor [HTML]{DFDDDD} 0.93 & 0.03 & 0.02 & 0.01\\
    BP$_3^{nm}$ & 0.50 & 0.32 & 0.03 & \cellcolor [HTML]{DFDDDD} 0.60 & 0.08 \\\hline\hline
    
  \end{tabular}
  \caption{{Selection criteria which have been used to define model independent search regions with jet multiplicities two and four are shown. The aplanarity 
  variable is defined  as $A = \frac{3}{2} \lambda_3$, where $\lambda_3$ is the smallest eigen value of the normalized momentum tensor of the jets (see Ref.~\cite{Bjorken:1969wi} for detail). The model independent 95\% CL upper limits derived by the ATLAS \cite{ATLAS-CONF-2019-040} on 
  the visible BSM contributions to the multijet cross-sections ($\langle \epsilon \sigma\rangle_{obs}^{95}$) for the above signal regions are also provided. The predictions for {\em mUED} and {\em nmUED} scenarios for three selected benchmark points (BPs), listed in Table~\ref{mUEDBP} and Table~\ref{nmUEDBP}, respectively, are also presented.}}
  \label{SRs}
\end{table}

\subsection{Validation}
\begin{table}[t!]
  \centering
  \begin{tabular}{|c|c|c|}
    \hline\hline
    \multirow{2}{2em}{Process} &\multicolumn{2}{c|}{SUSY gluino pair production}\\
    &\multicolumn{2}{c|}{$m_{\tilde{g}}=2200$ GeV and $m_{\tilde{\chi}^{0}_{1}}=600$ GeV} \\\hline
    \multirow{2}{3em}{Cuts}  & \multicolumn{2}{c|}{Absolute efficiency in$\%$ }\\\cline{2-3}
    & Appendix B of Ref.~\cite{ATLAS-CONF-2019-040} & Our Simulation \\
    \hline\hline
    preselection+$\rm{n}_j\geq$2 & 100.0 & 99.9\\
    $\rm{n}_j\geq$4 &92.9 &93.7 \\
    $\Delta\phi(j_{1,2,(3)},p_T^{\rm mis})_{min}>0.4$ &77.6 &74.7 \\ 
    $\Delta\phi(j_{i>3},p_T^{\rm mis})_{min}>0.2$ &69.1 &64.0 \\  
    $p_T(j_4)>100$ [GeV] &61.3 &55.7 \\
    $|\eta(j_{1,..,4})|<2.0$ &55.7 &50.2 \\
    $\rm{Aplanarity}>0.04$ &38.7 &33.5 \\
    $\frac{E_T\!\!\!\!\!\!/~}{\sqrt{H_T}}>16 [\sqrt{\rm{GeV}}]$&24.1 &17.9 \\
    $\rm{m}_{\rm{eff}}>1000$[GeV] &24.1 &17.9 \\
    \hline\hline
  \end{tabular}
  \caption{Cut flow table for signal region SR4j-1.0. {The cut efficiencies in the} second column {are} provided by the ATLAS {collaboration in Ref.~}\cite{ATLAS-CONF-2019-040}. {Cut efficiencies resulting from our simulation are presented in the third column for comparison.}}\label{table_validation}
\end{table}
Since we are following the ATLAS multijet analysis, validation of our analysis against the ATLAS results is very important.
{In Table~17 of Ref.~\cite{ATLAS-CONF-2019-040}, ATLAS collaboration has presented cut-flow table for their simulated gluino pair production events at $\sqrt s=$13 TeV for} gluino mass $m_{\tilde{g}} = 2200$ GeV and
the lightest neutralino (the spin-half supersymmetric partners of the SM EW bosons) mass $m_{\tilde{\chi}^{0}_{1}} = 600$ GeV. {For validation purpose,} we have also generated {gluino pairs} up to two extra partons {in \texttt{MG5\_aMC@NLO} \cite{Alwall:2014hca} with the {\bf NNPDF23LO} \cite{Ball:2014uwa} parton distribution functions. Subsequent decays, showering and hadronization are simulated in \texttt{Pythia 8.2} \cite{Sjostrand:2007gs,Sjostrand:2014zea}.} The CKKW-L merging scheme \cite{Catani_2001} is employed for matching and merging. Hadronized events are passed
into \texttt{Delphes 3} \cite{deFavereau:2013fsa} for object reconstructions and the implementation of cuts. The cut efficiencies supplied by the ATLAS are presented alongside ours in Table~\ref{table_validation}. The excellent agreement between
the ATLAS analysis and our simulation bolsters our confidence on our method to apply the same for other BSM scenarios like, {\em mUED} and {\em nmUED}. We must mention that similar exercises have been performed for other signal regions as well and numbers are consistent. However, we do not intend to present them here, but to move on to the actual goal of our study.
%

\section{{\em mUED} after LHC Run-II data}
\label{mued_model}

In this section, the {\em mUED} model is put to test under the lens of LHC data collected at 13 TeV. As mentioned in the introduction, {\em mUED} has one extra spatial dimension ($y$) compactified on a circle of radius $R$ which signifies the length scale under probe at the LHC. {At the tree-level,} the masses of KK states for a given KK level are almost degenerate leaving little space for the decay 
products to get registered at the {LHC} detector. {In {\em mUED},} radiative corrections to the masses play a very crucial role to remove the degeneracy. Loop corrections to the KK masses in an orbifolded theory are logarithmically divergent. Since, {\em mUED} is an effective theory which remains valid up to certain cut-off scale {($\Lambda$)}, the radiative corrections are proportional to the logarithm of {$\Lambda$}~\cite{Cheng:2002iz, Hooper:2007qk}. Therefore, {the phenomenology of} {\em mUED} is completely specified 
by only two parameters: the compactification radius\footnote{{The inverse of radius of compactification ($R^{-1}$) determines the overall mass scale of KK particles for a given KK level and hence, determines the production cross-sections of KK particles at the LHC.}} ($R$) and the cut-off scale\footnote{{The cut-off scale ($\Lambda$) controls the mass splitting between different KK particles for a given KK level and hence, determines the kinematics of {\em mUED} signatures at the colliders. The perturbativity of the $U(1)$ gauge coupling 
requires that $\Lambda < 40 R^{-1}$. It has been argued in Ref.~\cite{Datta:2012db} that a much stronger bound arises from the running of the Higgs-boson self-coupling and the stability of the electroweak vacuum. However, the results of Ref.~\cite{Datta:2012db} relies on the lowest-order calculations and the inclusion of higher-loops can substantially change these results. Therefore, in our analysis, we varied $\Lambda$ in the range 3--40$R^{-1}$.}} ($\Lambda$). {Low energy observables like muon $g-2$ \cite{Nath:1999aa,Agashe:2001ra}, 
flavor changing neutral currents \cite{Chakraverty:2002qk,Buras:2003mk,Agashe:2001xt}, $Z \to b\bar{b}$ decay \cite{Oliver:2002up}, the $\rho$-parameter \cite{Appelquist:2002wb}, $\bar{B} \to X_s \gamma$ \cite{Haisch:2007vb} and other electroweak precision tests \cite{Rizzo:1999br,Strumia:1999jm,Carone:1999nz,Gogoladze:2006br} put a lower bound of about 300-600 GeV\footnote{{KK parity ensures that one-loop {\em mUED} corrections to all electroweak observables are cut-off
independent \cite{Dey:2004gb,Dey:2003yh} and thus, the low energy constraints on $R^{-1}$ are almost independent of $\Lambda$. The observables start showing cut-off sensitivity of various degrees as one goes beyond one-loop or considers more than one extra dimension.}} on $R^{-1}$. On the other hand, consistency with WMAP/PLANCK--measured \cite{Komatsu:2010fb,Ade:2015xua} DM relic density data puts an upper bound of about 1.4 TeV \cite{Hooper:2007qk} on $R^{-1}$. Given this upper limit, it is extremely plausible that experiments at the LHC can either discover or rule out {\em mUED} which will be the key focus of discussion in the following.}

\begin{table}
  \begin{center}
   \begin{tabular}{|c||c|c||c|c|c|c|c|c||}
     \hline\hline
BPs & $R^{-1}$ & $\Lambda R$ & $m_{G^{(1)}}$ & $m_{Q^{(1)}}$ & $m_{q^{(1)}}$& $m_{W^{(1)}/Z^{(1)}}$& $m_{L^{(1)}}$& $m_{B^{(1)}}$ \\
    & [TeV]& & [TeV]& [TeV]& [TeV]& [TeV]& [TeV]& [TeV]  \\\hline\hline
BP$^m_1$ & 2.0 & 3 &  2.222 &  2.143 &  2.124 &  2.045 &  2.022 &  1.998 \\
BP$^m_2$ & 1.75 & 40 &  2.341 &  2.171 &  2.122 &  1.877 &  1.814 & 1.748 \\
BP$^m_3$ & 1.4 & 30 &  1.840 & 1.668 & 1.710 & 1.495 & 1.448 & 1.399 \\\hline\hline
\end{tabular}
 \caption{{\em mUED} benchmark points and mass spectra of relevant level-1 KK particles. }
 \label{mUEDBP}
 \end{center}
 \end{table}

{In order to discuss the production, decay, and the resulting collider signatures of the KK particles, and to present the numerical results, we have chosen three benchmark points (BPs) listed in Table~\ref{mUEDBP} along with the masses of relevant level-1 KK particles.} {Being strongly interacting, the} level-1 KK gluons $G^{(1)}$ and singlet ($q^{(1)}$) as well as doublet ($Q^{(1)}$) KK quarks  {are expected 
to be copiously pair produced at the LHC at 13 TeV center-of-mass energy. These level-1 KK particles subsequently decay to the SM particles and the LKP via on/off-shell lighter intermediate KK particles. {It is important to mention that in the framework of {\em mUED} scenario, the mass hierarchies among different level-1 KK particles are determined by the radiative corrections
only and hence, are independent of $R^{-1}$ and $\Lambda$. As a result, the decay branching ratios of the level-1 KK particles are also practically independent of $R^{-1}$ and $\Lambda$. As the spectra in Table~\ref{mUEDBP} suggest, $G^{(1)}$, being the heaviest among the level-1 KK particles,  can decay to both singlet ($q^{(1)}$) and doublet ($Q^{(1)}$) quarks with 
almost\footnote{{There is a slight kinematic preference for $G^{(1)}~\to~q q^{(1)}$ decay.}} same branching ratios. A singlet level-1 KK quark ($q^{(1)}$) decays only to $B^{(1)}$ in association with a SM quark. Similarly, a doublet level-1 KK quark ($Q^{(1)}$) decays preferably to $W^{(1)\pm}$ or $Z^{(1)}$ accompanied by a SM quark.
Mass spectra in Table~\ref{mUEDBP} shows that the hadronic decays of the $W^{(1)\pm}$ are closed kinematically. Therefore, it decays to all three level-1 KK doublet lepton flavors universally (both $L^{(1)\pm} \nu$ and $L^{(0)\pm} \nu^{(1)}$). Similarly, $Z^{(1)}$ can decay only to $L^{(1)\pm} l^{\mp}$ or $\nu^{(1)}{\nu}$ (with branching fractions being determined by the corresponding SM couplings).
The level-1 KK leptons finally decay to $B^{(1)}$ and an ordinary (SM) lepton. In all three BPs (in {\em mUED} in general), $B^{(1)}$ is the lightest KK particle {\it i.e.} the LKP. Therefore, the production and subsequent decays of level-1 KK quarks/gluons at the LHC give rise to a} final state consisting of a number of jets and/or leptons and missing transverse momentum. {However, the small mass splittings
between level-1 KK $W^{\pm}/Z$ and leptons as well as level-1 KK leptons and the LKP (see Table~\ref{mUEDBP}) would render very soft leptons in the final state. Thus, we concentrate only on the hadronic final states to probe the parameter space of
{\em mUED} at the LHC at $\sqrt s=13$ TeV and 139 fb$^{-1}$ integrated luminosity data as per the ATLAS strategy~\cite{ATLAS-CONF-2019-040}.} {Pair productions of level-1 KK quarks/gluons are simulated in} \texttt{MG5\_aMC@NLO} \cite{Alwall:2014hca} event-generator.  {The subsequent decays}, {initial state radiation} (ISR), {final state radiation} (FSR), hadronization
{\em etc.} {are simulated in} \texttt{Pythia 8.2}. For the purpose of reconstruction and analysis of the events, we designed our own analysis code with very close proximity to the ATLAS utilized {object reconstruction} criteria and selection cuts. 

\begin{figure}[t]
\begin{center}
\includegraphics[width=0.8\linewidth]{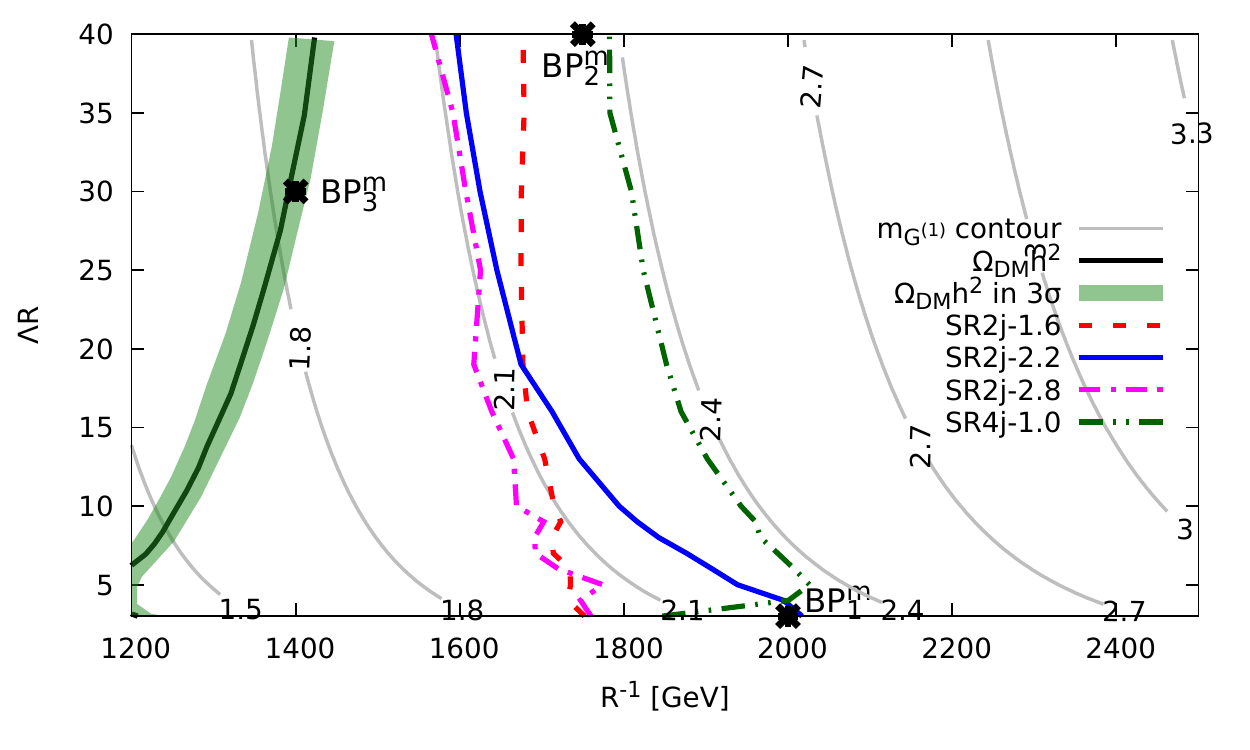}\\
 \caption{95\% {CL} exclusion plot on $R^{-1}$-$\Lambda R$ plane from different SRs {(see Table~\ref{SRs})} of {13 TeV} ATLAS {search \cite{ATLAS-CONF-2019-040} for multijets $ + E_T\!\!\!\!\!\!/~$ with} 139 fb$^{-1}$ integrated luminosity {data}. The region left to {exclusion} lines {corresponding to different SRs} are {ruled out} at 95\% {CL}. Level-1 KK gluon mass ($m_{G^{(1)}}$) contours are laid over
 as grey lines along with corresponding masses printed in TeV. {The three benchmark points, 
 listed in Table~\ref{mUEDBP}, are also shown in filled black dots.} The black solid line with green band of 3$\sigma$ significance surrounding it represents the region that give correct dark matter relic density \cite{Cornell:2014jza}. The entire region right to the relic density ($\Omega_{\rm{DM}} \rm{h}^2$) curve  is said to be ruled out in view of over-closure of the universe.}
\label{mued-bound}
\end{center}
\end{figure}

The results are summarized in {Table~\ref{SRs} and} Fig.~\ref{mued-bound}. {We present the final exclusion bound in Fig.~\ref{mued-bound}  on the {\em mUED} parameter space for each of the SRs listed in Table~\ref{SRs}. The region in the $R^{-1}$--$\Lambda R$ plane to the left of a given exclusion curve is ruled out at 95\% CL. Fig.~\ref{mued-bound} also shows level-1 KK gluon mass (in TeV) contours. For large $ \Lambda R$, the strongest bound comes from 4-jet final state (in particular,
SR4j-1.0 signal region) which excludes level-1 KK gluon mass below about 2.37 TeV.} Note that the parameter space with lower $\Lambda R \lesssim 5$ seems somewhat less restricted. {For small $\Lambda R$, the strongest bound of about 2.22 TeV on level-1 KK gluon mass results from SR2j-2.2. }
{The numerical predictions for signal multijet $+~E_T\!\!\!\!\!/~$ cross-sections in different SRs are presented in Table~\ref{SRs} for the {\em mUED} benchmark points defined in Table~\ref{mUEDBP}.  BP$_1^m$ represents the part of {\em mUED} parameter space characterized by small $\Lambda R \sim 3$ and hence, a highly degenerate mass spectra for level-1 KK 
particles. As a result, the decays of level-1 quarks/gluons give rise to very soft jets at the LHC. For such a scenario, a mono-jet like final state comprising a single high $p_T$ jet, resulting primarily from initial state radiation, accompanied by missing transverse energy is a promising channel. Table~\ref{SRs} shows that BP$_1^m$ is excluded from SR2j-2.2 which is indeed a mono-jet like \cite{Aad:2015zva,Aaboud:2017phn} signal region. BP$_2^m$(BP$_3^m$) corresponds to 
large $\Lambda R \sim 40(30)$ and hence, relatively larger mass splittings between level-1 KK particles. At the parton level, the pair (associated) production of level-1 KK gluons (in association with level-1 KK quarks) and their subsequent decays give rise to four (three) hard jets. Additional jets also arise from initial
state radiation. Therefore, for large $\Lambda R$ regions, four-jet channels (in particular, SR4j-1.0) are the most promising ones for estimating the bound as can be seen from Table~\ref{SRs} as well as from Fig.~\ref{mued-bound}.}

Although, we do not claim to have performed {any} dark matter related analysis, for the sake of completeness, we have shown the {relic density} bound {on $R^{-1}$-$\Lambda R$ plane} from {Ref.~\cite{Cornell:2014jza}}. Potential reason for its inclusion is that the bound from dark matter abundance appears to be the most severe one and strips off a large chunk of parameter space. In Fig.~\ref{mued-bound} the narrow green strip centered around solid black line shows the parameter region with correct dark matter relic density. The band signifies the 3$\sigma$ tolerance level. {The parameter space} on the left of the {RD} line results  into relic densities which are smaller than the RD observed by WMAP/PLANCK. Therefore, this region is allowed in the sense that one {can always} concoct scenarios
{with multi-component dark matter} in order to evade such strict constrain from relic abundance. {However, the entire region right to the relic density
curve in Fig.~\ref{mued-bound}  corresponds to relic densities larger than the WMAP/PLANCK result and hence, is ruled out. Therefore, we can conclude from Fig.~\ref{mued-bound} that the region of $R^{-1}$-$\Lambda R$ plane, which is consistent with WMAP/PLANCK RD data, has already been ruled out by the ATLAS multijet $+~E_T\!\!\!\!\!/~$~searches at 13 TeV LHC with
139 fb$^{-1}$ integrated luminosity.} Hence, we shift our focus on the {\em non-minimal UED (nmUED)} where an enhanced number of parameters offer rich phenomenology. Next section is slotted for discussion on the theoretical set up of the model.

\section{{\em nmUED} : Model description}\label{nmUED_model}
 
The assumption of vanishing boundary terms in {\em mUED} is somewhat unnatural, since they can anyway be generated {at} the loop-level. Moreover, these  boundary-localized terms (BLT) obey all the symmetries of the model~\cite{Flacke:2014jwa}. The {\em non-minimal} version of the model  ({\em nmUED}) takes these BLTs into account. Every boundary-localized term is associated with a parameter, which we generally denote by $r$. {The presence of these} unknown {BLT} parameters drastically alters the {\em nmUED} mass-spectrum compared to the {\em mUED} one. {Moreover,} the interaction vertices of {involving} various {non-zero} KK{-modes are non-trivially modified by a multiplicative factor known as,} overlap integrals. However, before going into the collider phenomenology of {\em nmUED} scenario, we briefly introduce the theoretical set-up
of {\em nmUED} scenario.}

    The {most general {\em nmUED}} action is {required to be} invariant under the gauge symmetry of the {SM {\em i.e.,} invariant under} $SU(3)_C \times SU(2)_{W} \times U(1)_Y${, as well as the Lorentz symmetry in 4D,} {and can be written as,}
    \begin{equation}
\mathcal{S}_{\text{\em nmUED}}~=~\mathcal{S}_{\text{gluon}}+\mathcal{S}_{W}+\mathcal{S}_{B}+\mathcal{S}_{{\rm quark}}+\mathcal{S}_{{\rm lepton}}+ {\cal S}_{\text{scalar}},
    \end{equation}
{where, the individual parts of the full {\em nmUED} action, $\mathcal{S}_{\text{\em nmUED}}$, are discussed in the following.} {The gauge part of the action is given by,}
\begin{align}
&\mathcal{S}_{\text{gluon}} = \int d^4 x \int_{0}^{\pi R} dy \Bigg\{ - \frac{1}{4} G^{a}_{MN} G^{aMN} +
                 \Big(\delta(y) + \delta(y-\pi R) \Big) \Big[ - \frac{r_G}{4} G^{a}_{\mu\nu} G^{a\mu\nu} \Big] \Bigg\},& \nonumber \\
&\mathcal{S}_{W} = \int d^4 x \int_{0}^{\pi R} dy \Bigg\{ - \frac{1}{4} W^{i}_{MN} W^{iMN} +
                 \Big(\delta(y) + \delta(y-\pi R) \Big) \Big[ - \frac{r_W}{4} W^{i}_{\mu\nu} W^{i\mu\nu} \Big] \Bigg\},& \nonumber \\
&\mathcal{S}_{B} = \int d^4 x \int_{0}^{\pi R} dy \Bigg\{ - \frac{1}{4} B_{MN} B^{MN} + 
		  \Big(\delta(y) + \delta(y-\pi R) \Big) \Big[ - \frac{r_B}{4} B_{\mu\nu} B^{\mu\nu} \Big] \Bigg\}.
		 \label{gauge} &
\end{align}
where, $G^{a}_{MN}, W^{i}_{MN}, B_{MN}$ stand for 5D field-strength tensors corresponding to the $SU(3)_C$, $SU(2)_{W}~\text{and}~U(1)_Y$ gauge fields, respectively. The symbols $M,~N$ run for $0,1,2,3,5$ and the Greek indices run for $0,1,2,3$. The actions clearly consist of two parts. {The} first {parts are} the usual gauge kinetic term in 5D. The second part{s} {are} the brane-(also called boundary) localized kinetic terms (BLKTs). These terms appear only at the boundaries of the brane and the bulk, as can be seen by the presence of delta functions.
We consider boundary parameters at the {two} orbifold-fixed points to be the same which ensures the conservation of KK parity. Now we briefly describe {the} fermionic {parts of the action which can be written as,}
\begin{eqnarray}
\label{5d_quark}
\mathcal{S}_{{\rm quark}} &=& \sum_{j=1}^3\int d^4 x \int_{0}^{\pi R} dy \Big[\overline{Q}_j i\Gamma^{M} \mathcal{D}_{M} Q_j + r_Q \{ \delta(y) + \delta(y-\pi R) \} \overline{Q}_j i\gamma^{\mu} \mathcal{D}_{\mu} P_L Q_j \nonumber \\ 
 & & + \overline{U}_j i\Gamma^{M} \mathcal{D}_{M} U_j + r_U \{ \delta(y) + \delta(y-\pi R) \} \overline{U}_j i\gamma^{\mu} \mathcal{D}_{\mu} P_R U_j \nonumber \\ 
 & & + \overline{D}_j i\Gamma^{M} \mathcal{D}_{M} D_j + r_Q \{ \delta(y) + \delta(y-\pi R) \} \overline{D}_j i\gamma^{\mu} \mathcal{D}_{\mu} P_R D_j\Big],\\
\label{5d_lepton}
\mathcal{S}_{{\rm lepton}} &=& \sum_{j=1}^3\int d^4 x \int_{0}^{\pi R} dy \Big[\overline{L}_j i\Gamma^{M} \mathcal{D}_{M} L_j + r_L \{ \delta(y) + \delta(y-\pi R) \} \overline{L}_j i\gamma^{\mu} \mathcal{D}_{\mu} P_L L_j \nonumber \\
 & & + \overline{E}_j i\Gamma^{M} \mathcal{D}_{M} E_j + r_L \{ \delta(y) + \delta(y-\pi R) \} \overline{E}_j i\gamma^{\mu} \mathcal{D}_{\mu} P_R E_j \Big], 
\end{eqnarray}
where, 5D {quark (lepton) doublet and singlets are denoted by $Q~(L)$ and $U/D~(E)$, respectively,  $j=1,2,3$ is the generation index, $\Gamma_M = (\gamma_{\mu},i\gamma_5)$ denotes $\gamma$-matrices in 5D and $\mathcal{D}_{M}$ is the gauge covariant derivative in 5D.}
{Finally,} the action corresponding to the {5D Higgs} field is given by,
\begin{align}
 {\cal S}_{\text{scalar}} &= \int d^4 x \int_{0}^{\pi R} dy  \Bigg\{ ({\cal D}^M \Phi)^{\dagger}({\cal D}_M \Phi) + \mu^2_5 \Phi^{\dagger}\Phi -
 \lambda_5 (\Phi^{\dagger}\Phi)^2  \nonumber \\
 &+ \{ \delta(y) + \delta(y+\pi R) \} 
  \left( r_{\Phi} ({\cal D}^{\mu} \Phi)^{\dagger}({\cal D}_{\mu} \Phi)  +  \mu^2_B \Phi^{\dagger}\Phi - \lambda_B (\Phi^{\dagger}\Phi)^2 \right)\Bigg\},
  \label{scalar}
\end{align}
where, $\Phi$ is {the 5D} Higgs. $\mu_5$ and $\lambda_5$ represent the 5D bulk
Higgs mass parameter and scalar self-coupling, respectively. The BLKT parameter for scalar field is
$r_{\Phi}$; $\mu_B$ and $\lambda_B$ are the boundary-localized Higgs mass parameter and the scalar quartic coupling, respectively.
We must mention that all the BLT parameters ($r_i$ where $i$ stands for $G$, $W$, $B$, $Q$, $L$ and $\Phi$ fields) are dimensionful parameters. However, we express our results in section \ref{nmued_collider} in terms of scaled BLT parameters $R_i = r_i R^{-1}$ as is customary.

 The {\em nmUED} action written in the previous paragraph contains
the information of the full theory in 5D.
5D fields can be expanded into $x_{\mu}$ and $y$ dependent
 parts where $x_{\mu}$ is the usual 4D space-time coordinates and $y$ is the extra dimension coordinate which is compactified on a $S_1/Z_2$ orbifold.
 Once the mode expansions are fed into the actions and the extra dimensional coordinate, $y$, is integrated out, we obtain a 4D effective theory involving
 the SM particles as well as their KK modes. For example, the mode expansions for the 5D gluon can be written as,
\begin{align*}
  G^{a(n)}_{\mu}(x,y)~= \sum^{\infty}_{n=0} G^{a(n)}_{\mu}(x)~f^{(n)}_G(y),~{\rm with~~}
  f^{(n)}_{G}(y) &= N_{G^{(n)}} \times
\begin{cases}
\displaystyle \frac{\cos(m_{G^{(n)}} y)}{C_{G}}  & \text{for $n$ even }\\
\displaystyle -\frac{\sin(m_{G^{(n)}} y)}{S_{G}}  & \text{for $n$ odd }
\end{cases},
\end{align*}
{where, $C_{G} =\cos \left(m_{G^{(n)}}\pi R/2\right)$ and $S_{G} =\sin \left(m_{G^{(n)}}\pi R/2\right)$.
} {Note that the above expansion together with the} boundary conditions {give rise to}
 the following transcendental equations:
\begin{align}
 r_G m_{G^{(n)}} =
\begin{cases}
\label{KKmass1}
- 2\tan \left({\frac{m_{G^{(n)}} \pi R}{2}}\right) & \text{for $n$ even} \\
2\cot \left({\frac{m_{G^{(n)}}\pi R}{2}}\right) & \text{for $n$ odd}
\end{cases}
\end{align}
{In the framework of {\em nmUED}, the mass of the $n$-th level KK gluon ($m_{G^{(n)}}$) is obtained by solving} these transcendental equations. The normalization of the wave function {($N_{G^{(n)}}$) is given by,} 
%
\begin{equation} \label{eq_norm}
N_{G^{(n)}} = \left[\left(\frac{\pi R}{2}\right) \left(1 + \frac{r_G^2 m_{G^{(n)}}^{2}}{4} + \frac{r_G}{\pi R}\right)\right]^{-\frac{1}{2}}.
\end{equation}
Such KK decomposition and transcendental equations are common for all the 5D fields. {Therefore, in {\em nmUED} scenario, the masses for the KK modes of other SM particles are also given by the solution of transcendental equations similar to those in Eq.~\ref{KKmass1} with appropriate BLT parameters.}
%
%

{It is interesting to note that the phenomenology of the level-1 electroweak gauge sector of {\em nmUED} is significantly different from that of {\em mUED} since the masses and mixings of the level-1 KK EW gauge bosons in {\em nmUED} non-trivially depend on the BLT parameters $r_W$, $r_B$ and $r_{\Phi}$. In the context of {\em mUED}, the masses of the lightest (i.e. the LKP which is the DM candidate in the theory)
and next-to-lightest level-1 KK gauge boson are determined by the radiative corrections. In addition, the extent of mixing between the level-1 $U(1)_Y$ and $SU(2)_W$ KK gauge bosons is minuscule, unless $R^{-1}$ is very small. Therefore, in {\em mUED}, the lightest and next-to-lightest level-1 KK gauge bosons are, for all practical purposes, essentially the level-1 excitations of $U(1)_Y$ and $SU(2)_W$ gauge bosons, respectively. However,
in presence of the various overlap integrals involving the gauge and scalar BLT parameters, 
the mixing between the level-1 $U(1)_Y$ and $SU(2)_W$ gauge bosons could be large in the framework of {\em nmUED}. Moreover, depending on the choice of $r_W$, $r_B$ and $r_{\Phi}$, the LKP in {\em nmUED} could be either a level-1 excitation of $U(1)_Y$ gauge boson, or a level-1 excitation
of $SU(2)_W$ gauge boson. These facts, in turn, have profound implications for the dark matter phenomenology. {Note that in {\em mUED}, due to little freedom available for determining the mass spectrum and mixing, the observed value of dark matter RD provides a stringent upper bound on $R^{-1}$ which essentially rules out
the model at the LHC.} However, the additional parameters $r_W$, $r_B$ and $r_{\Phi}$ in {\em nmUED} play a crucial role to lift the RD upper bound on $R^{-1}$. It has been shown in Ref.~\cite{Flacke:2017xsv} that with proper choice of $r_W$ and $r_B$, larger values of $R^{-1}$ is possible without
conflicting with the measured value of dark matter RD. {The freedom in setting the mass spectrum of level-1 KK particles at required value also helps specific co-annihilation
channels to contribute more and thus, obtain the required RD.} We intend to address these issues related to the dark matter phenomenology of {\em nmUED} in a future article. In the present article, we focused on the 
collider bounds on the masses of level-1 KK quarks and gluons in the framework of {\em nmUED}.}

{Before going into the collider phenomenology of {\em nmUED}, it is important to mention that the couplings, involving the zero-mode and non-zero mode KK particles, are also modified non-trivially by factors known as, the overlap integrals.}
These {coupling} modifications appear once we plug in the KK expansion{s} in 5D Lagrangian and integrate {over} the extra dimensional {coordinate, $y$}. {Note that a} generic {interaction of a level-$l$} gauge boson ($V^{(l)}$) {with
a pair of level-$m$ and $k$ fermion-antifermion ($\Psi^{(m)}(x)$ and $\bar \Psi^{(k)}(x)$)  results from the following term in the 5D action after compactifying the extra dimensional coordinate $y$:}
\begin{align*}
{\cal S}_{int}&=~\tilde{g}~ \int d^4x \int dy~ \sum_{k,l,m} \left(\bar \Psi^{(k)}(x)\; f_{\Psi}^{(k)}(y)  \right) \times
                 \gamma^\mu \left( V_{\mu}^{(l)}(x)\;f_{V}^{(l)}(y) \right) \times 
                 \left( \Psi^{(m)}(x)\; f_{\Psi}^{(m)}(y) \right)\\
                 &=~ \sum_{k,l,m} \int d^4x~\tilde{g}\left[\int dy~f_{\Psi}^{(k)}(y)f_{V}^{(l)}(y)f_{\Psi}^{(m)}(y)\right]\times \left[\bar \Psi^{(k)}(x) \gamma^\mu V_{\mu}^{(l)}(x)\Psi^{(m)}(x)\right],
\end{align*}
{where, $\tilde g$ is the corresponding gauge coupling in 5D.} The connection between 5D gauge coupling $\tilde{g}$ and its 4D counterpart is given by $g = \tilde{g}/\sqrt{r_V + \pi~R}$, where $r_V$ is the corresponding BLT {parameter} for the {5D} gauge boson $V_M$.
{Note that the integration over the extra dimensional coordinate, $y$, is non-zero only for certain combinations of the KK numbers $(k,l,m)$ and hence, acts as a selection rule known as, KK number conservation, for interactions involving different KK level particles.}  In {\em mUED}, the integration over the extra dimensional
coordinate in the above equation is either one (for KK number conserving interactions) or zero (for KK number violating interactions) depending on the choice of $(k,l,m)$. However, the presence of BLTs {in {\em nmUED} result
into a} $y$ profile of KK excitations {which is} different from the {\em mUED} case {and hence,} gives rise to non-trivial overlap integral. 
  \begin{figure}[h]
  \begin{center}
  \includegraphics[width=0.7\columnwidth]{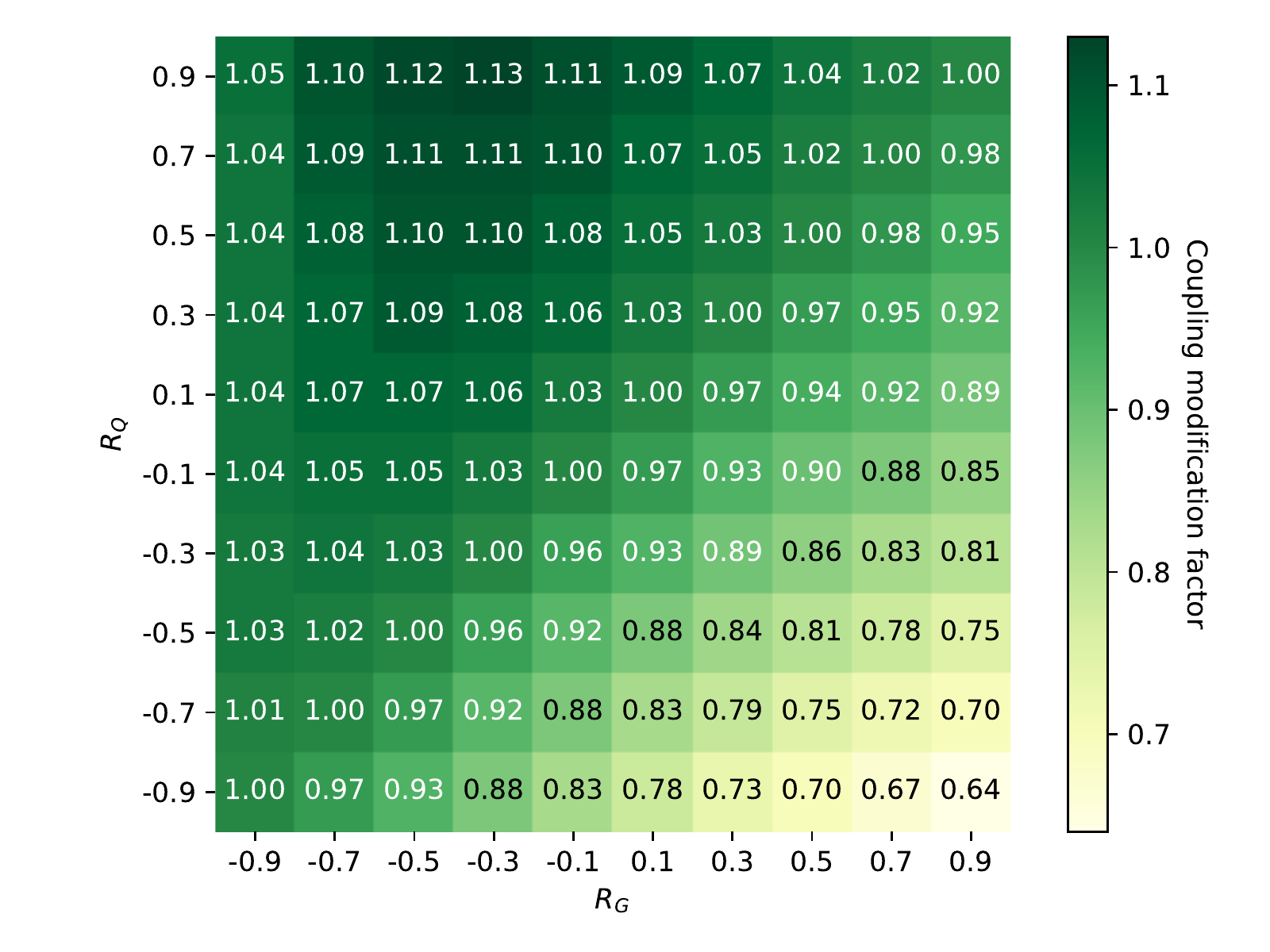}
  \end{center}
  \caption{The $Q^{(0)}$ $G^{(1)}$ $Q^{(1)}$ coupling {modification} factors {are} plotted against scaled BLT parameters $R_G$ and $R_Q$ corresponding to gluon and quark, respectively.
  The actual coupling for the $Q^{(0)}$ $G^{(1)}$ $Q^{(1)}$ vertex is given by these factors multiplied by the QCD coupling $g_s$.}
  \label{gprime-rQ-rG}
  \end{figure}
{Depending on the values of the BLT parameters, the overlap integrals} can enhance or reduce a particular coupling and thereby, influence the phenomenology of the model. {In Fig.~\ref{gprime-rQ-rG}}, 
the modification factors {for the gauge coupling involving a level-1 KK gluon, a level-1 KK quark and a SM quark} have been plotted against the gluon and quark scaled
\footnote{{{From now on, we display our results in terms of dimensionless scaled BLT parameters defined as $R_{i} = r_{i}/R$ as defined earlier.}}} BLT parameters. {Fig.~\ref{gprime-rQ-rG} shows significant deviation
from unity in different parts of $R_G$--$R_Q$ plane. {It can be deduced from Fig.~\ref{gprime-rQ-rG} } that in certain parts of $R_G$--$R_Q$ plane, one could obtain an enhancement (suppression)
as large as 13\%~(36\%) in the interaction strength of $Q^{(0)}$$G^{(1)}$$Q^{(1)}$ vertex compared to the interaction strength of pure QCD vertex.}

\subsection{Collider Phenomenology}\label{nmued_collider}
\begin{figure}[h]
  \begin{center}
  \includegraphics[width=0.495\columnwidth]{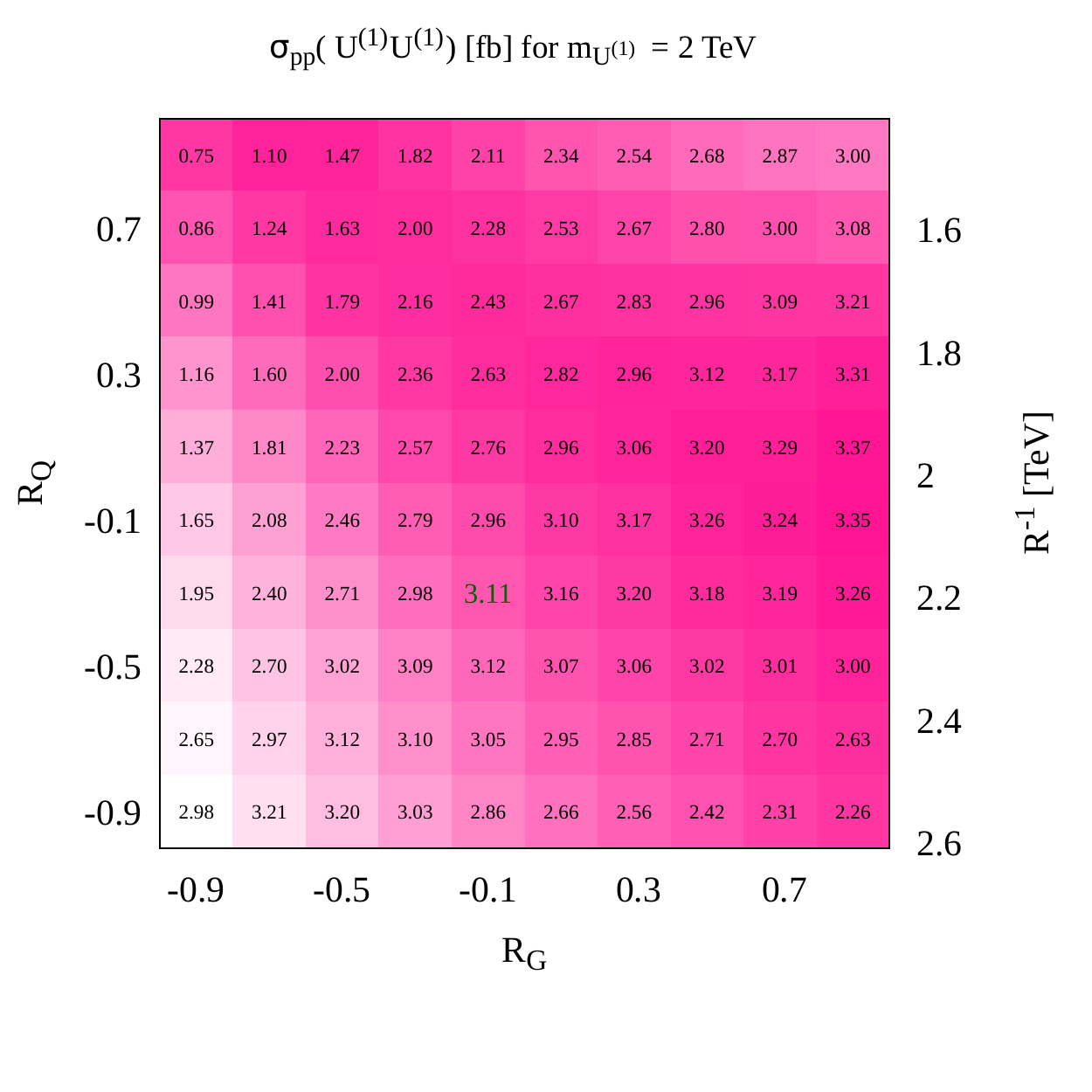}
  \includegraphics[width=0.495\columnwidth]{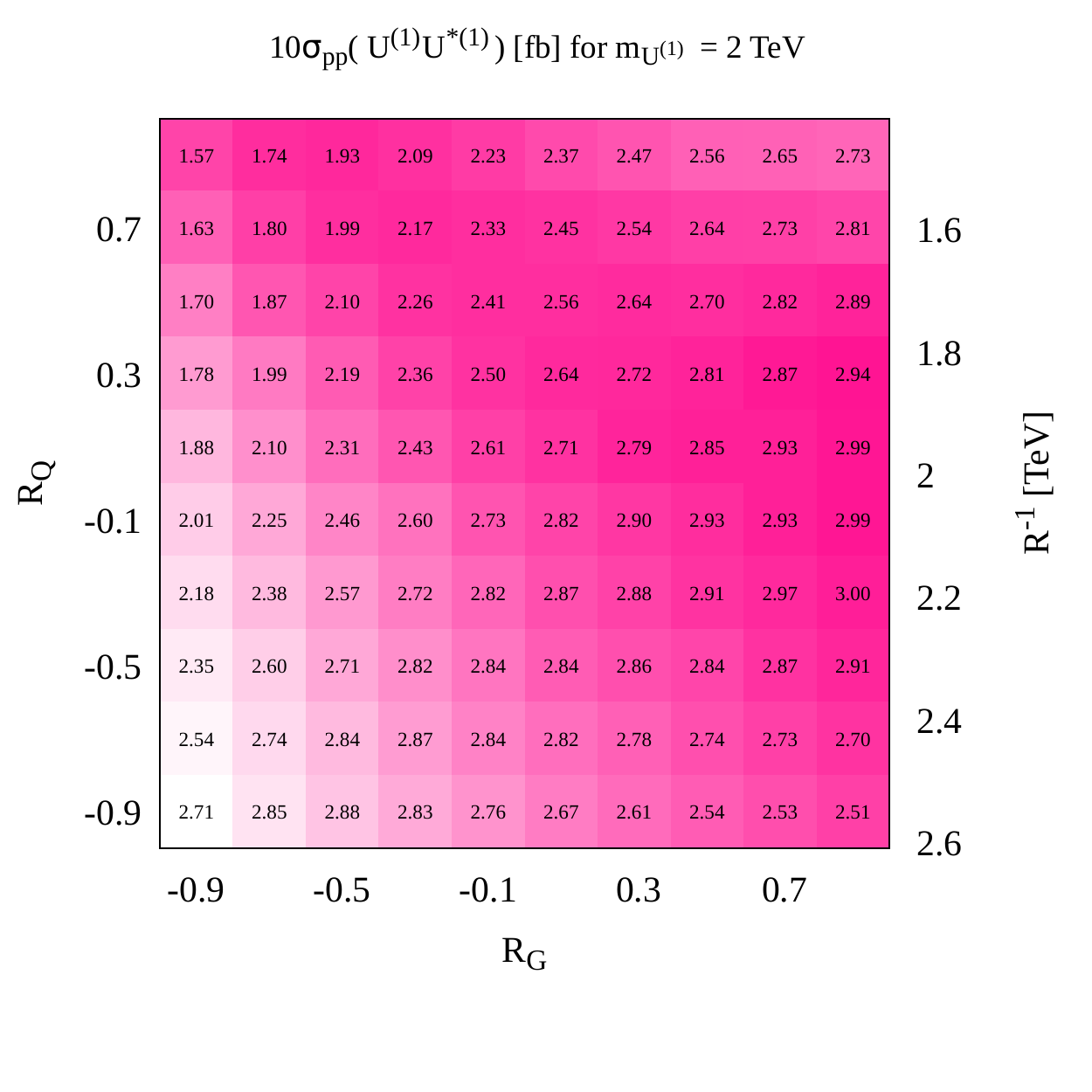}
  \end{center}
  \caption{Cross sections (in fb) for the up-type level-1 quark-quark (left panel)  and quark-antiquark (right panel) pair productions are presented on $R_G$--$R_Q$ plane for the LHC 
  at 13 TeV center of mass energy. Instead of fixing $R^{-1}$, the level-1 KK quark mass is kept fixed at $m_{U^{(1)}}=2$ TeV. The $y_2$-axis shows the variation of $R^{-1}$.}
  \label{prod_dudu}
  \end{figure}

After discussing the {\em nmUED} model, mass spectrum and coupling modifications, we are now equipped enough to study its collider phenomenology and impose bounds from the ATLAS search for multijet plus $E_T\!\!\!\!\!\!/~$ final states. However, before delving into the ATLAS analysis,
{it is important to} discuss {the} productions {of different KK particles and their subsequent} decay{s} {in the framework of {\em nmUED}.} {The LHC being a proton-proton collider, we only consider the QCD pair productions of level-1 KK quarks/gluons in our analysis. Unlike {\em mUED}\footnote{{In the framework of {\em mUED}, the pair and associated production cross-sections of level-1 KK quarks and KK gluons depend only on the masses $m_{Q^{(1)}}$ and $m_{G^{(1)}}$ and hence, on $R^{-1}$ and $\Lambda$.}}, the {\em nmUED}
QCD pair production cross-sections of level-1 KK particles are determined by radius of compactification as well as the BLT parameters for the quarks and gluons. } {The inverse of radius of compactification sets the overall {mass} scale {for the level-1 KK particles} in {\em nmUED}}, over which $R_G$ and $R_Q$ fix the masses of the KK gluons and KK quarks, respectively.
In addition, the BLT parameters also govern the  {strength of interactions involving the SM and level-1 KK particles which are crucial for the productions as well as decays of the level-1 KK particles. For example, the $Q^{(0)}$$G^{(1)}$$Q^{(1)}$ coupling (the dependence of which on $R_G$ and $R_Q$ is shown in Fig.~\ref{gprime-rQ-rG}) 
appears in all the relevant QCD production of level-1 KK particles at the LHC namely, $G^{(1)}G^{(1)},~G^{(1)}Q^{(1)}/q^{(1)},~Q^{(1)}Q^{(1)},~Q^{(1)}\bar Q^{(1)},~q^{(1)}q^{(1)},~q^{(1)}\bar q^{(1)}$ etc.} {To illustrate the dependence of QCD productions of level-1 KK particles on $R_Q$ and $R_G$, we have presented the pair/associated production cross-sections of level-1 KK gluon and pair productions of level-1 KK quarks in Fig.~\ref{prod_g1g1} and \ref{prod_dudu}, 
  \begin{figure}[t]
  \begin{center}
  \includegraphics[width=0.495\linewidth]{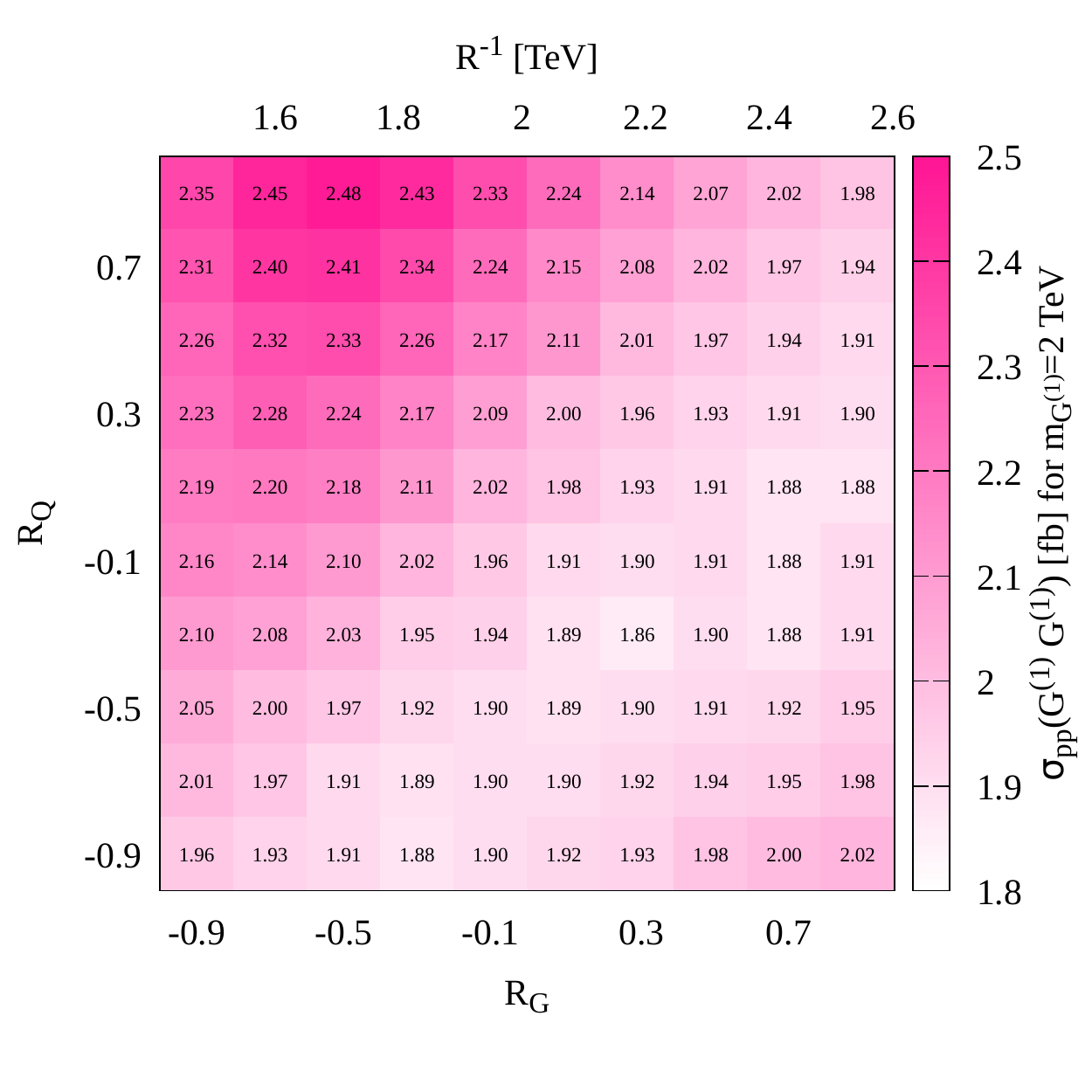}
  \includegraphics[width=0.495\linewidth]{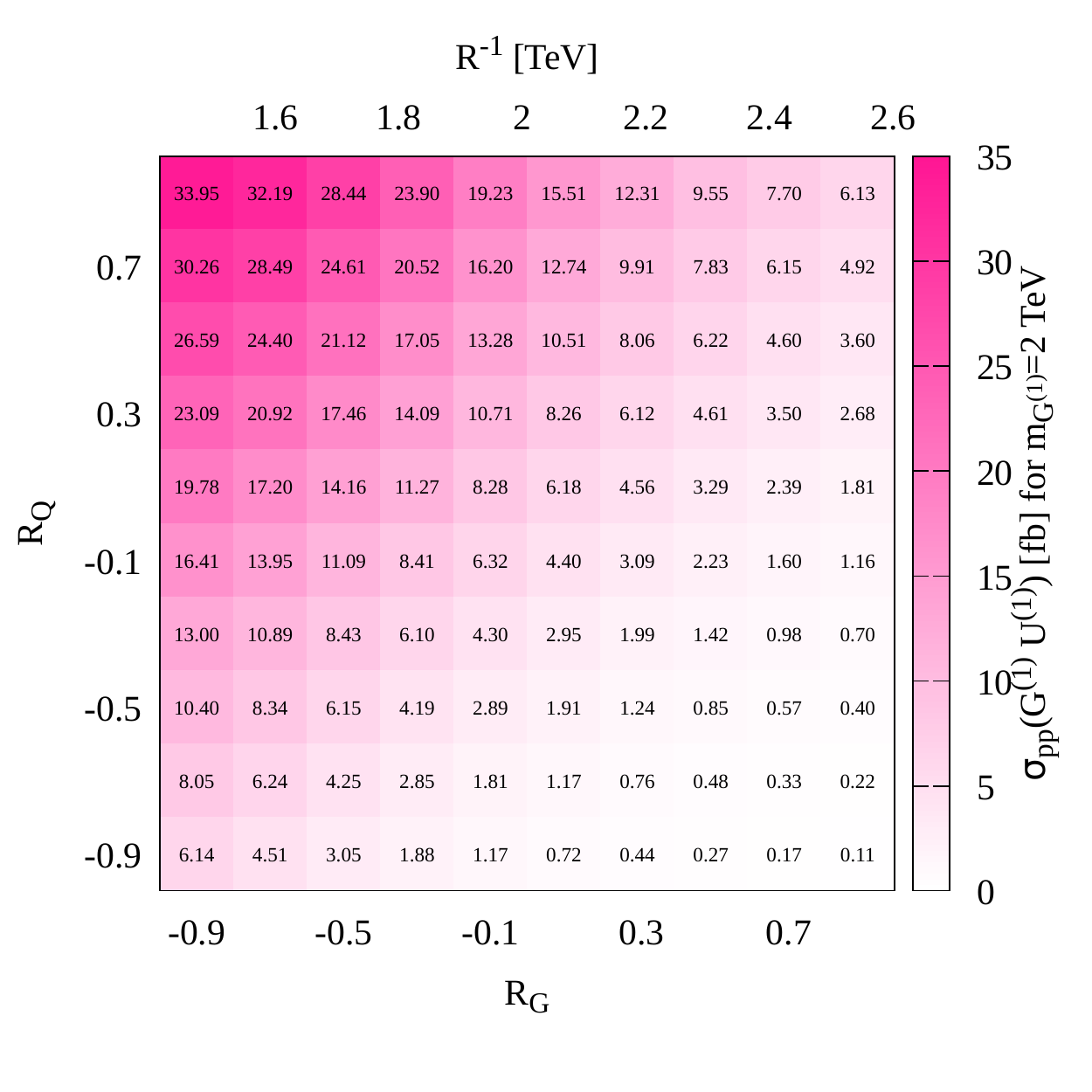}
  \end{center}
  \caption{Cross sections (in fb) for the pair (left panel)  and associated (right panel) production level-1 KK gluon are presented on $R_G$--$R_Q$ plane for the LHC at 13 TeV center of mass energy. Instead of fixing $R^{-1}$, the level-1 KK gluon mass is kept fixed at $m_{G^{(1)}}~=~2$ TeV. The $x_2$-axis shows the values of $R^{-1}$.}
  \label{prod_g1g1}
  \end{figure}
respectively. The left panel of Fig.~\ref{prod_g1g1} shows the pair production cross-sections of level-1 KK gluons for fixed $m_{G^{(1)}}~=~2$ TeV\footnote{{In {\em nmUED}, the level-1 KK gluon mass is obtained by solving the transcendental equation in Eq.~\ref{KKmass1} and hence, $m_{G^{(1)}}$ depends on both $R^{-1}$ and $R_G$. For a given value of $R_G$, one can obtain $m_{G^{(1)}}~=~2$ TeV by suitably choosing a value for $R^{-1}$. Therefore, the plots in Fig.~\ref{prod_g1g1} (as well as the plots in Fig.~\ref{prod_dudu}) do not correspond to a particular value of $R^{-1}$. To clearly display the dependence of QCD cross-sections on $R_Q$ and $R_G$, one needs to minimize the dependence on parton densities and phase space factors and hence, ensures fixed values for the final state particle masses. This motivates us to present the cross-sections in Fig.~\ref{prod_g1g1}(\ref{prod_dudu}) for a fixed $m_{G^{(1)}}(m_{Q^{(1)}})$ instead of a fixed $R^{-1}$.}}. The dominant contribution\footnote{{Quark-antiquark initiated diagrams also contribute to $\sigma(pp\to G^{(1)}G^{(1)})$. Gluon densities being larger than quark/antiquark densities at the LHC energies, the quark-antiquark initiated contributions to $\sigma(pp\to G^{(1)}G^{(1)})$ are suppressed compared to the gluon-fusion contribution.}} to the $G^{(1)}G^{(1)}$ production ($\sigma(pp\to G^{(1)}G^{(1)})$) at the LHC results from the gluon-gluon initiated process with a level-1 KK gluon in the $t(u)$-channel. The vertices involved in the Feynman diagrams of $gg\to G^{(1)}G^{(1)}$ are purely QCD vertices which do not get modified and hence, 
$\sigma(gg\to G^{(1)}G^{(1)})$ depends only on $m_{G^{(1)}}$. However, some of the Feynman diagrams (in particular, the $t (u)$
channel level-1 KK quark exchange diagrams) for the subdominant $q\bar q\to G^{(1)}G^{(1)}$ production channel involve $Q^{(0)}$$G^{(1)}$$Q^{(1)}$ vertex which gets modified. Therefore,
the variation of $\sigma(pp\to G^{(1)}G^{(1)})$ in Fig.~\ref{prod_g1g1} (left panel) results from the $R_G$ and $R_Q$ dependence of quark-antiquark initiated contribution to the total cross-section. 
An important fact is that the two Feynman diagrams namely, the $s$-channel gluon exchange diagram and $t(u)$-channel $Q^{(1)}$ exchange diagram, contributing to the quark-antiquark initiated production of $G^{(1)}$-pairs,
interfere destructively. For a given $R_G$ (and hence, fixed $R^{-1}$) in Fig.~\ref{gprime-rQ-rG}, increasing $R_Q$ corresponds to decreasing $m_{Q^{(1)}}$ and hence, stronger destructive interference which tends to decrease the cross-section. On the other hand, the 
coupling modification factors increase with increasing $R_Q$ (see Fig.~\ref{gprime-rQ-rG}) which tends to increase the cross-section. These two competing factors explain the pattern of $\sigma(pp\to G^{(1)}G^{(1)})$ variation on $R_Q$ and $R_G$ as displayed in Fig.~\ref{prod_g1g1} (left panel). 

The associated production cross-section ($\sigma(pp\to G^{(1)}U^{(1)})$) of a level-1 KK gluon in association with a level-1 up-type KK quark\footnote{{At the LHC, the electroweak productions of level-1 KK quarks/gluons are negligible compared to the QCD productions. Therefore, we have neglected EW productions of level-1 quarks/gluons in our analysis.
Note that QCD cannot distinguish between u-type and d-type or singlet and doublet KK quarks. Therefore, the cross-sections presented here are valid for both singlet and doublet up-type quarks. Results will be different for level-1 down-type quarks due to the parton density of $d(\bar d)$-quark.}} as a function of $R_G$ and $R_Q$ is presented in Fig.~\ref{prod_g1g1} (right panel) for fixed $G^{(1)}$ 
mass of 2 TeV. Here, the mass of the level-1 up-type KK quark, however, is not constant over the $R_G$--$R_Q$ plane.
At the LHC, the $G^{(1)}U^{(1)}$ associated production is a quark-gluon initiated process which proceeds via the exchange of a level-1 KK quark or KK gluon in the $t (u)$-channel. Although, both the Feynman diagrams contributing to $\sigma(pp\to G^{(1)}U^{(1)})$ 
contain coupling which depends on BLT parameters, the large variation of $\sigma(pp\to G^{(1)}U^{(1)})$ in Fig.~\ref{prod_g1g1} (right panel) mainly occurs due to the variation of the level-1 KK quark mass in the final state. The $R_G$--$R_Q$ dependence of the production cross-sections of level-1 KK quark-quark pair 
($\sigma(pp\to U^{(1)}U^{(1)})$) and KK quark-antiquark pair ($\sigma(pp\to U^{(1)}\bar U^{(1)})$) are presented in the left panel and right panel, respectively, of Fig.~\ref{prod_dudu} for $m_{U^{(1)}} = 2$ TeV. In order to generate a fixed level-1 KK quark mass, $R^{-1}$ needs to be varied with $R_Q$. The variation of $R^{-1}$ is also depicted
as the $y_2$-axis in Fig.~\ref{prod_dudu}. The dominant contribution to $\sigma(pp\to U^{(1)}U^{(1)})$ comes from the quark-quark fusion process at the LHC through a level-1 KK gluon in the $t (u)$-channel.
The resulting variation of $\sigma(pp\to U^{(1)}U^{(1)})$  with respect to $R_G$--$R_Q$ is shown in Fig.~\ref{prod_dudu} (left panel). On the other hand, $\sigma(pp\to U^{(1)}\bar U^{(1)})$ receives contributions from quark-antiquark and gluon-gluon initiated processes. While the gluon-gluon initiated
channel depends only on $m_{U^{(1)}}$ and hence, independent of $R_G$ and $R_Q$ for a fixed $m_{U^{(1)}}=2$ TeV; the mild variation of $\sigma(pp\to U^{(1)}\bar U^{(1)})$ over $R_G$--$R_Q$ plane (see the right panel of Fig.~\ref{prod_dudu}) results from the sub-dominant quark-antiquark initiated process.}

{After discussing the productions of level-1 KK quarks/gluons, we will now discuss the decays of various level-1 KK particles and the resulting signatures at the LHC.}
Mass hierarchy {among} various KK particles {plays a crucial role in determining the} {the decay cascades of level-1 KK quarks/gluons and hence,} the {topology and} kinematics of the final state{s at the LHC}.
\begin{table}[h] 
  \begin{center}
   \begin{tabular}{|c||c|c||c|c|c|c|c|c||}
     \hline\hline
BPs & $R^{-1}$ & $(R_Q,R_G)$ & $m_{G^{(1)}}$ & $m_{Q^{(1)}}$ & $m_{W^{(1)}}$ &$m_{Z^{(1)}}$ &  $m_{L^{(1)}}$ & $m_{B^{(1)}}$ \\
    & [TeV]& &  [TeV] & [TeV]& [TeV]& [TeV]& [TeV] & [TeV] \\\hline\hline
BP$^{nm}_1$ &  1.9 & (-0.9,-0.1) &  1.963  &  2.559 &  1.913 & 1.914  &  1.906 & 1.900 \\
BP$^{nm}_2$ & 2.1 & (-0.1,-0.1) &  2.169 &  2.169 &  2.114 &   2.115 & 2.107 & 2.100 \\
BP$^{nm}_3$ &  2.0 & (-0.3,-0.7) &  2.531 &  2.209&   2.013 & 2.015 & 2.007 & 2.000\\\hline\hline
\end{tabular}
 \caption{{\em nmUED} benchmark points and mass spectra of relevant level-1 KK particles. }
 \label{nmUEDBP}
 \end{center}
 \end{table}
 {While, for {\em mUED}, the mass hierarchy among KK particles  of a given level is completely determined by the radiative corrections, the {\em nmUED} mass spectrum is determined by the BLT parameters which are free parameters of the theory. For example,} $R_G < R_Q$ would render a mass hierarchy {similar to {\em mUED}} {with} KK gluon  {being} more massive than KK quarks {while $R_G > R_Q$ results into KK quarks being heavier than the KK gluons.} {In order to discuss the decays, and the resulting collider signatures {\em nmUED} as well as present the numerical results, we have chosen three benchmark points (BPs) listed in Table~\ref{nmUEDBP} along with the masses of relevant level-1 KK particles. The BPs in Table~\ref{nmUEDBP} are characterized by $R^{-1}$ and $(R_Q,R_G)$. We have assumed fixed values for the BLT parameters\footnote{{It is important to mention the existing constraints on the BLT parameters. The absence of any tachyonic modes requires the scaled BLT parameters to be larger than $-\pi$. The presence of BLTs leads to the KK number violation at the tree level. The LHC searches for $Z^\prime$ in dilepton channels lead to stringent constraints on the production of level-2 gauge bosons via KK number violating couplings involving a level-2 gauge boson and a pair of SM fermions. In particular, $R_{W/B} > 0.2$ for $R^{-1} = 1.5 $ TeV ~\cite{Flacke:2017xsv} has already been excluded from the dilepton resonance searches at the LHC.}} in the EW sector namely, $R_W,~R_\Phi,~R_B$ and $R_L$. We consider $R_W=-0.02=R_{\Phi}$ and set $R_B$ to zero. This particular choice of the EW BLT parameters gives rise to 
 a LKP which is dominantly the level-1 KK excitation of the $U(1)_Y$ gauge boson ($B^{(1)}$) with significant mixing with the level-1 KK excitation of the neutral $SU(2)_L$ gauge boson ($W^{3(1)}$). Note that the EW level-1 KK gauge sector is highly degenerate (see Table~\ref{nmUEDBP}) with the dominantly $SU(2)_L$ level-1 KK gauge bosons namely, $W^{\pm(1)}$ and $Z^{(1)}$, being slightly heavier than the LKP. It has been shown in Ref.~\cite{Flacke:2017xsv} that such an EW level-1 sector of {\em nmUED} enhances the dark matter annihilation/co-annihilation cross-sections and hence, allows larger values of $R^{-1}$ without conflicting  with the WMAP/PLANCK measured value of the RD. We have also fixed the BLT parameters for leptons at $R_L = -0.01$ and scanned over negative\footnote{{For $R_B~=~0$, positive values of $R_Q$ or $R_G$ gives rise to a stable colored particle in the theory and hence, excluded.}} values of $R_Q$ and $R_G$.}

{For $R_G < R_Q$} with $m_{G^{(1)}} > m_{Q^{(1)}}$ (see BP$^{nm}_3$ in Table~\ref{nmUEDBP}), the level-1 KK gluon decays primarily to a SM quark and its level-1 KK counterpart. {Since the decays of level-1 KK quarks into a SM quark and $G^{(1)}$ are kinematically 
forbidden for BP$^{nm}_3$,} $m_{Q^{(1)}} > m_{W^{(1)}}/m_{B^{(1)}}$, usually result into a level-1 doublet KK quark decaying to a SM quark in association with a $W^{(1)\pm}$ / $Z^{(1)}$ / $B^{(1)}$. Note that for the  level-1 singlet KK quarks, the decay into $W^{(1)\pm}$ is highly suppressed. {In the scenario with $R_G > R_Q$ (see BP$^{nm}_1$ in Table~\ref{nmUEDBP}), the level-1 KK quarks, 
being heavier than $G^{(1)}$, dominantly decay into a SM quark in association with a level-1 KK gluon. On the contrary, $G^{(1)}$ undergoes a tree-level 3-body decay via an off-shell $Q^{(1)}$ into a SM quark-antiquark pair in association with a level-1 EW 
boson ($W^{(1)\pm}$/$Z^{(1)}$/$B^{(1)}$). 
  \begin{figure}[h]
  \begin{center}
  \includegraphics[width=0.49\columnwidth]{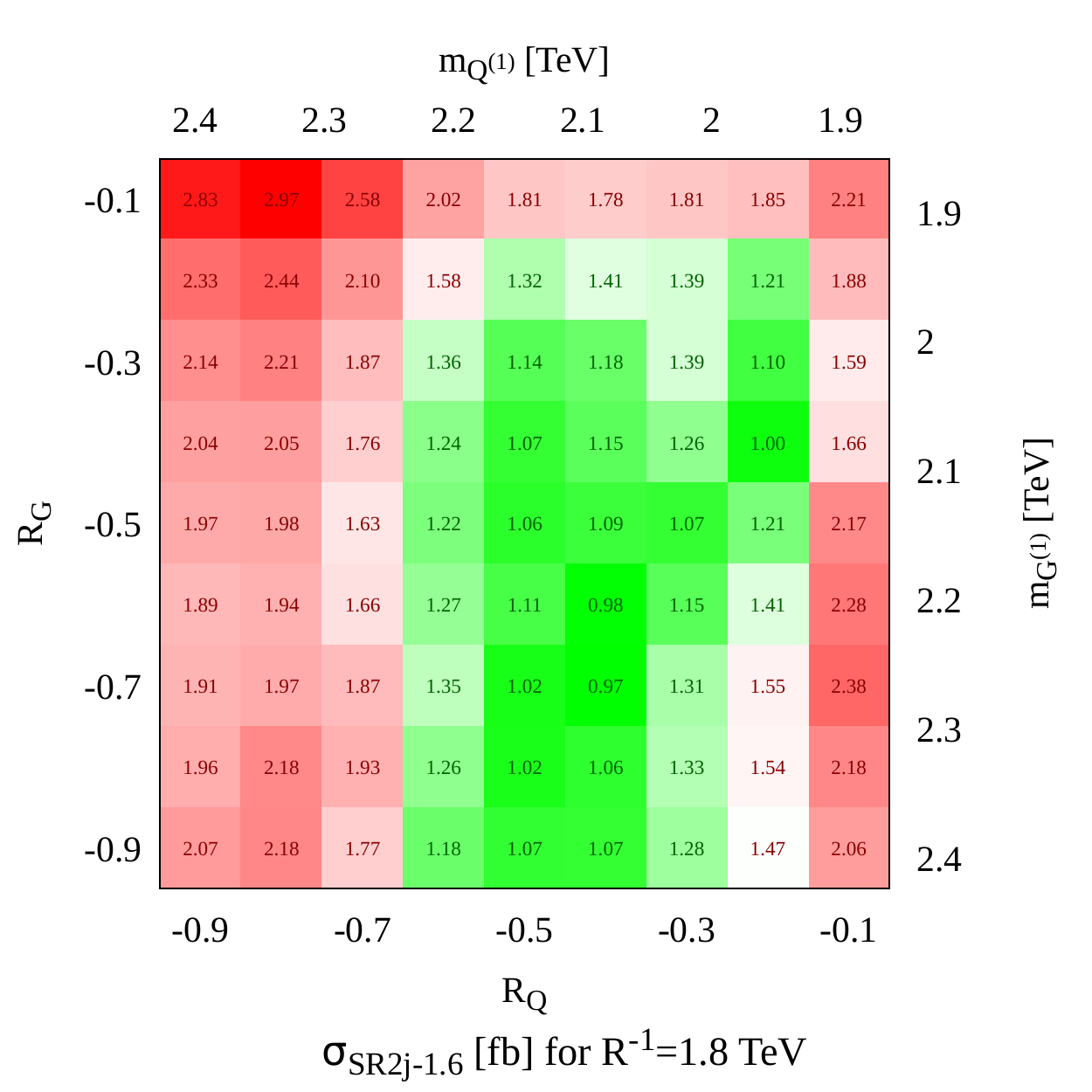}
  \includegraphics[width=0.49\columnwidth]{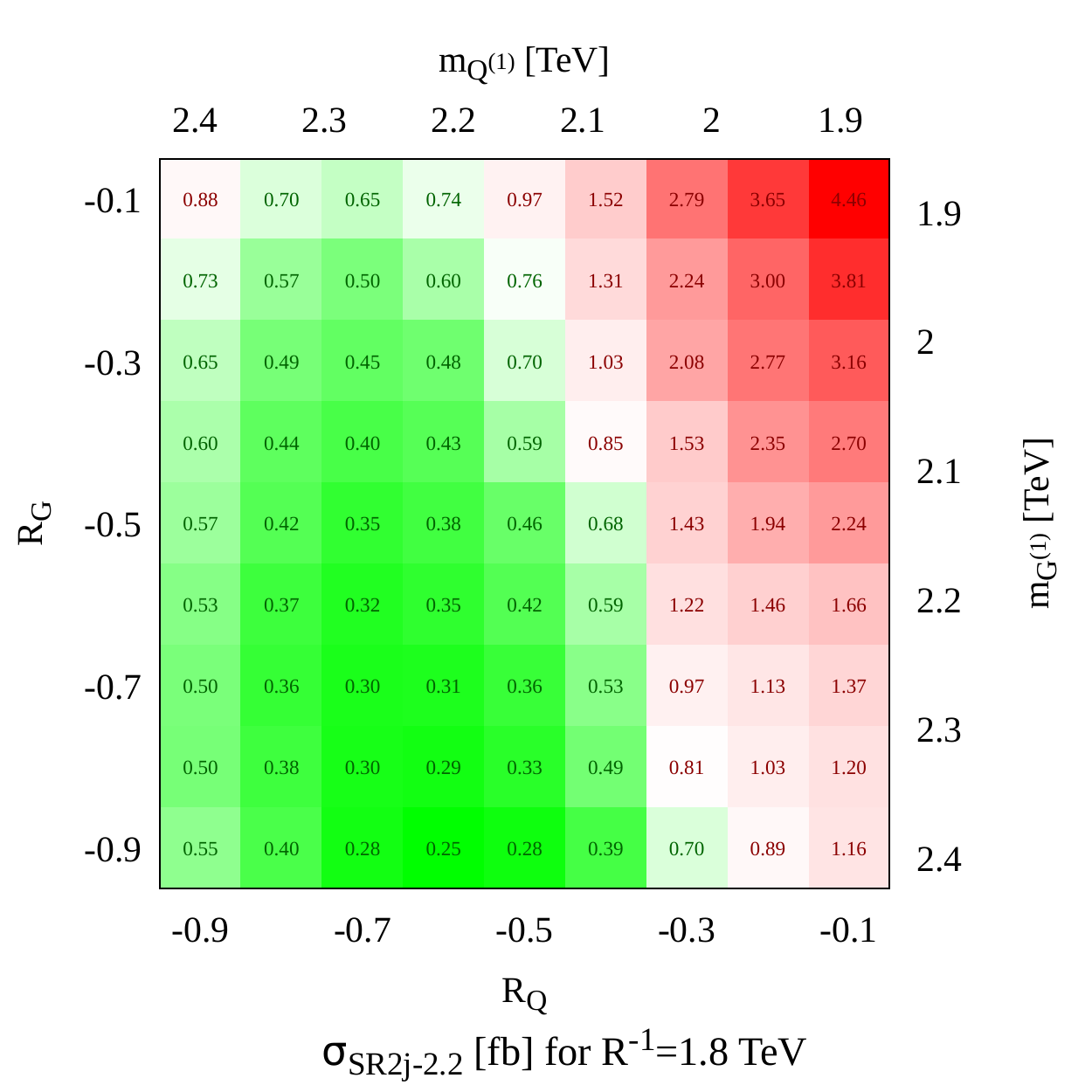}
  \includegraphics[width=0.49\columnwidth]{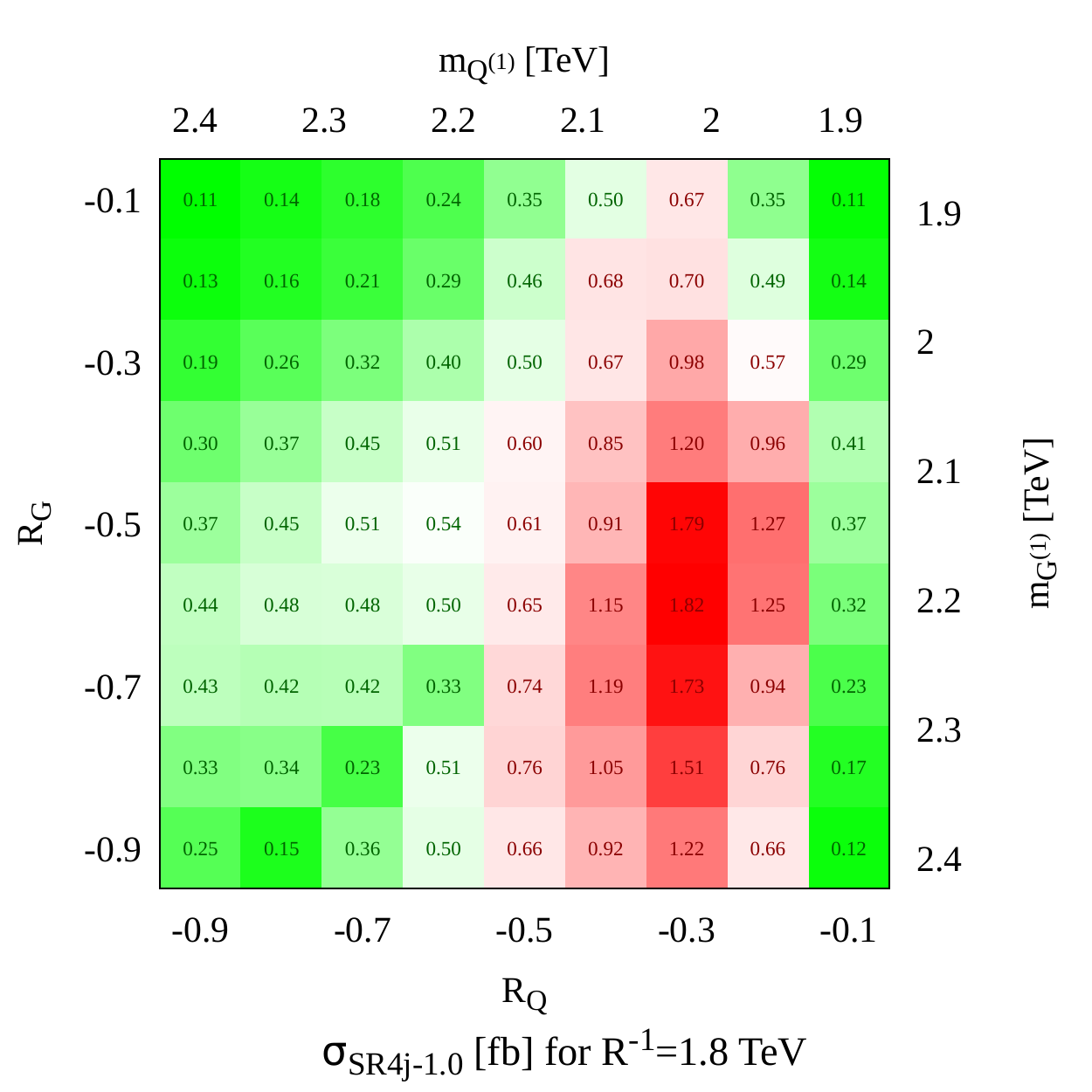}
  \end{center}
  \caption{Theoretical predictions for the visible cross-sections in SR2j-1.6 (top left panel), SR2j-2.2 (top right panel) and SR4j-1.0 (bottom panel) SRs are presented as a function of $R_Q$ ($x_1$-axis) and $R_G$ ($y_1$-axis) for {\em nmUED} scenario with $R^{-1}=1.8$ TeV, $R_{W(\Phi)}=-0.02$, $R_{L}=-0.01$ and $R_{B}=0.0$.
  The level-1 KK quark and KK gluon masses are also presented as $x_2$-axis and $y_2$-axis, respectively. For each of the signal regions, the excluded and allowed parts of parameter space are represented by red and green cells, respectively.}
  \label{exc1800}
  \end{figure}
The level-1 KK EW bosons subsequently decay into a pair of SM leptons and the LKP. The leptons arising from the decay of $W^{(1)\pm}$/$Z^{(1)}$ are usually very soft and hence, often remains invisible at the LHC detectors. Therefore, the pair/associated productions of level-1 KK quarks and KK gluons give rise to multijet in association with large missing transverse energy final states which will be discussed in the following in the context of recent ATLAS search \cite{ATLAS-CONF-2019-040} for multijet $+$ missing transverse energy final states at the LHC at 13 TeV center of mass energy and 139 fb$^{-1}$ integrated luminosity.}



%
  \begin{figure}[h]
  \begin{center}
  \includegraphics[width=0.49\columnwidth]{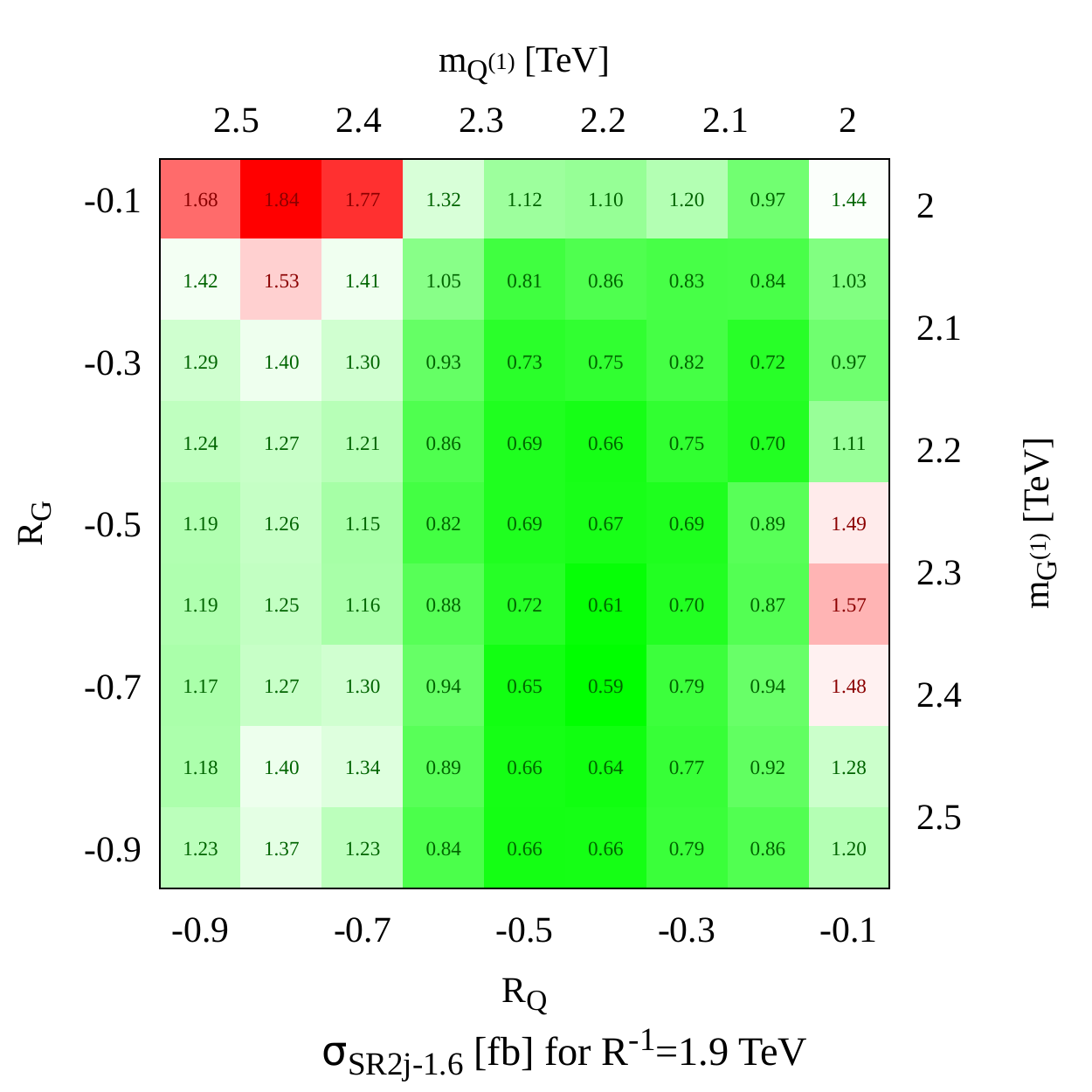}
  \includegraphics[width=0.49\columnwidth]{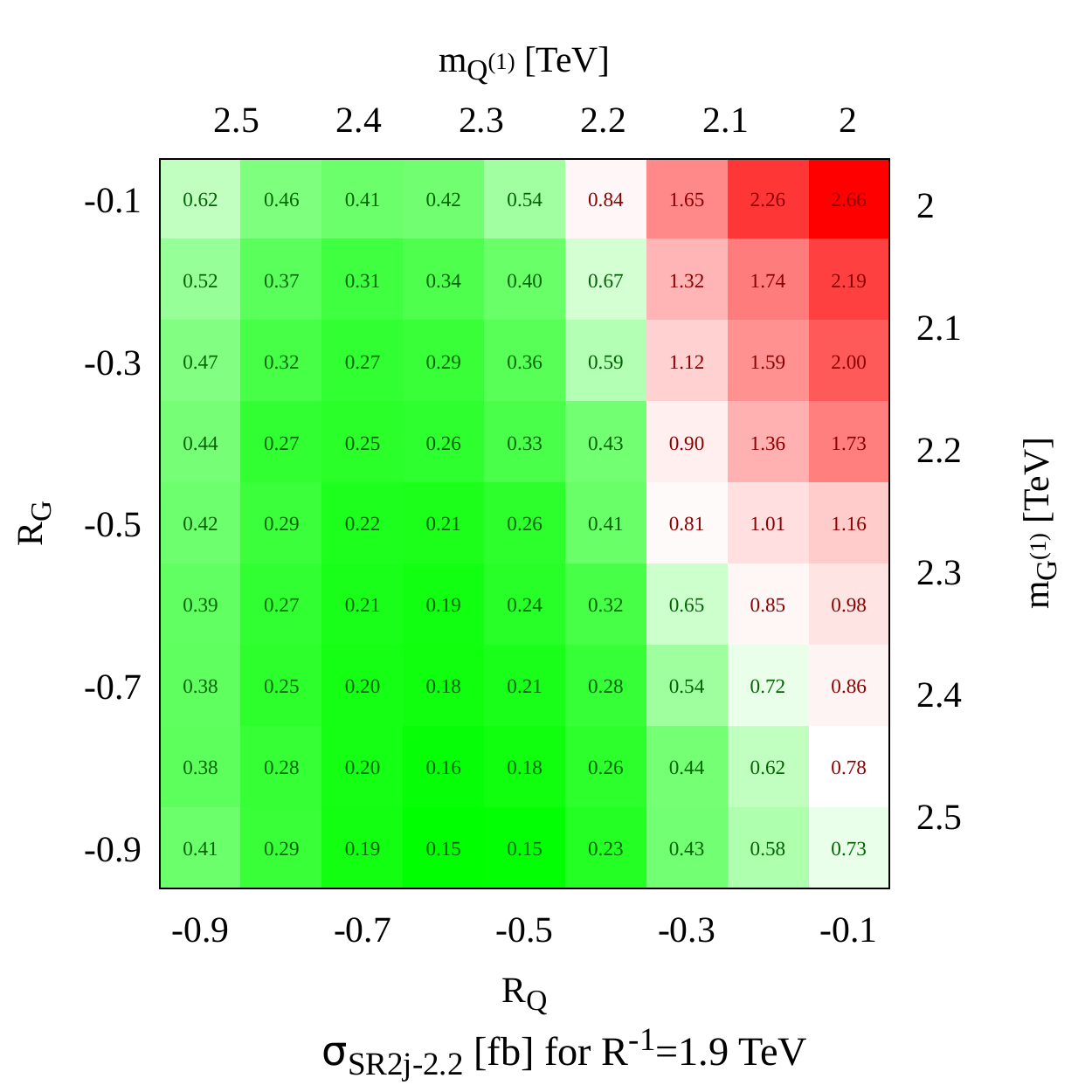}
  \includegraphics[width=0.49\columnwidth]{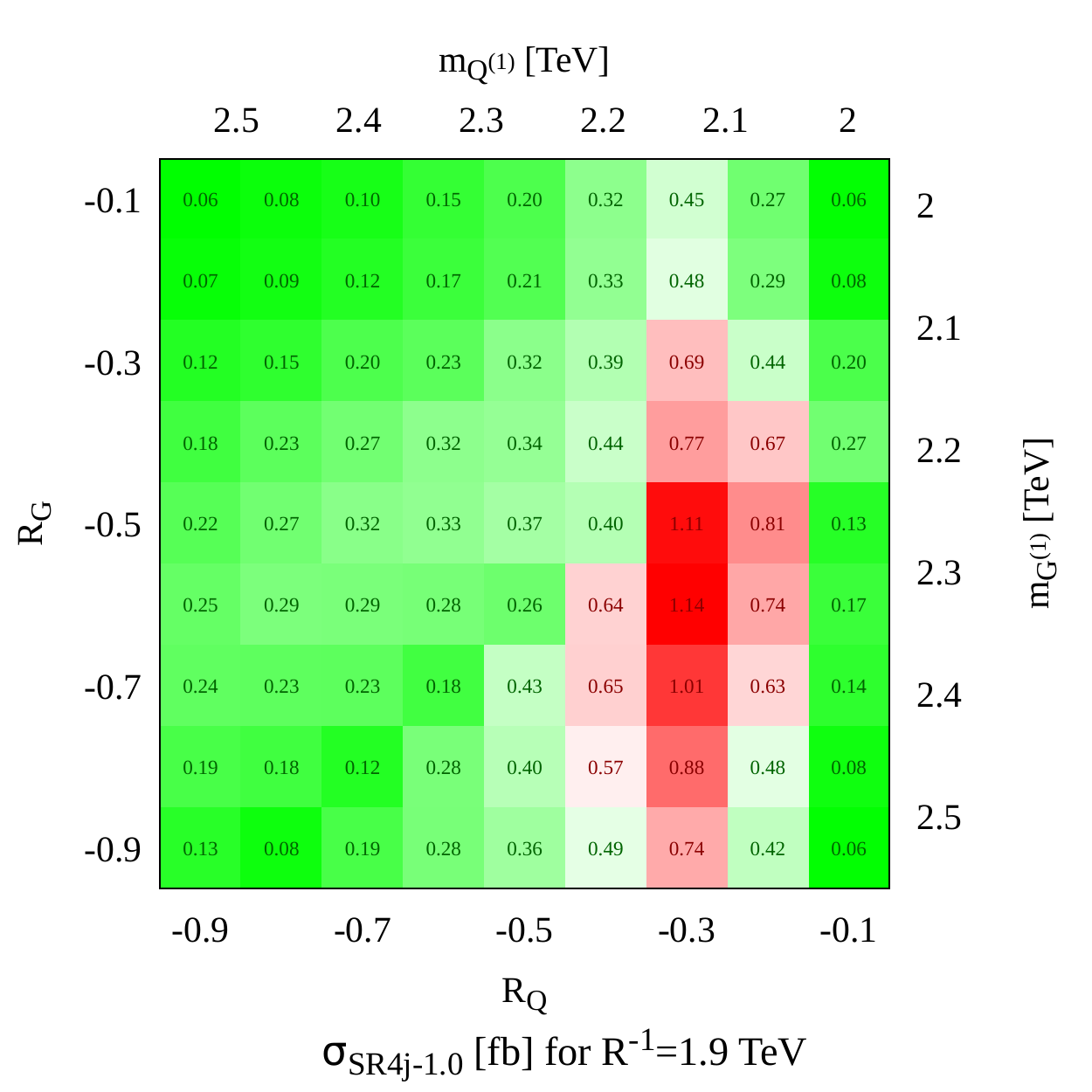}
  \end{center}
  \caption{Same as Fig.~\ref{exc1800} for $R^{-1}~=~1.9$ TeV.}
  \label{exc1900}
  \end{figure}


  \begin{figure}[t]
  \begin{center}
  \includegraphics[width=0.7\textwidth]{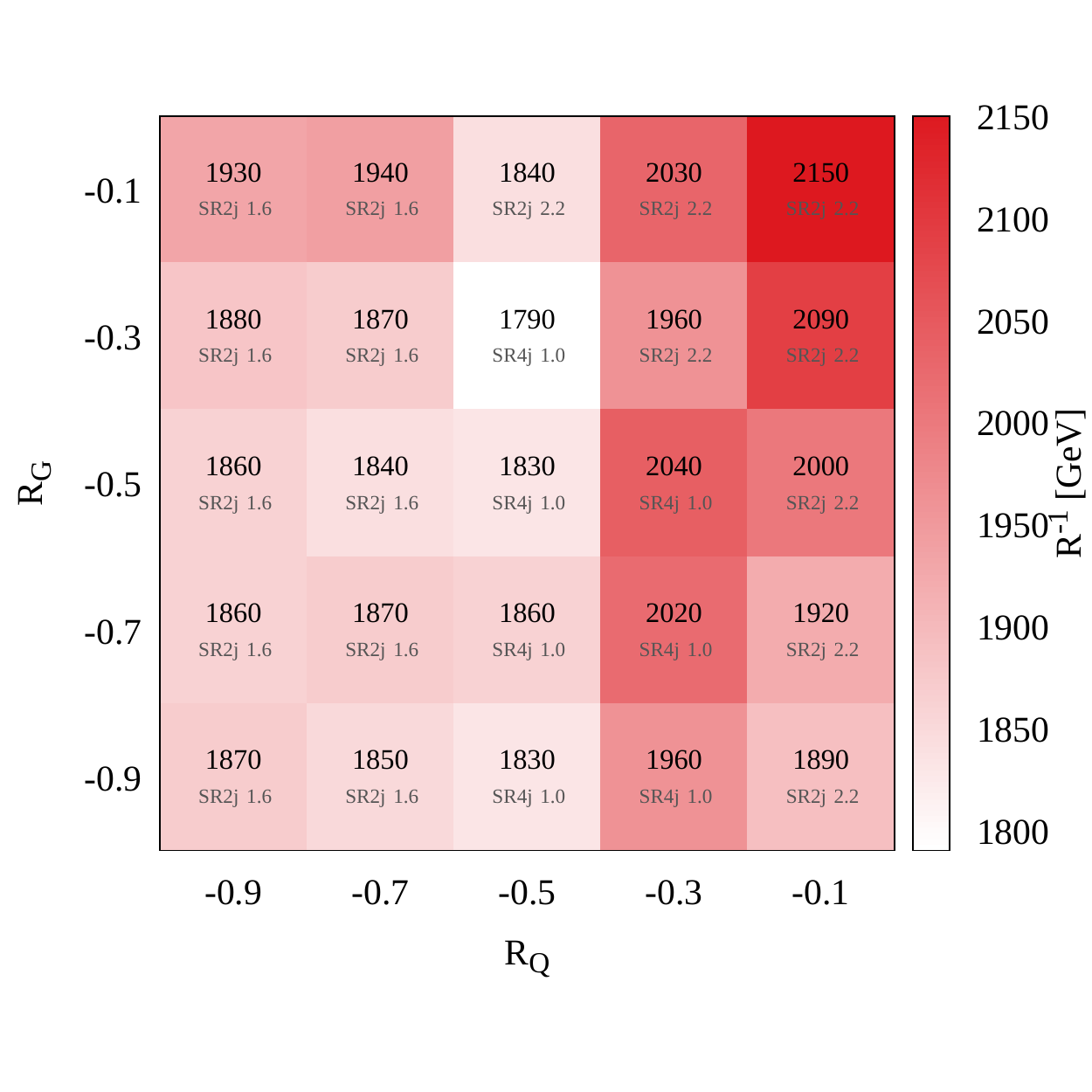}
  \caption{The lower bounds on $R^{-1}$ are presented as function of $R_Q$ and $R_G$. The signal region which leads to a particular lower bound on $R^{-1}$ for a given $R_Q$ and $R_G$ is also specified in the figure.}
    \label{excl_nmued}
    \end{center}
  \end{figure}

{The pair and associated productions of level-1 KK quarks and KK gluons are generated in MadGraph\footnote{{Note that {\em nmUED} UFO files, which are required for generating {\em nmUED} events in MadGraph, are not available in the literature. We have implemented the {\em nmUED} model in \texttt{FeynRules 2.0} \cite{Alloul:2013bka} to generate the UFO files.}}. 
The MadGraph generated events are  fed into \texttt{Pythia 8.2} for simulating decays, ISR, FSR and hadronization. We use our own analysis code for the object reconstructions and  computation of {\em nmUED} contributions to the multijet $+$ $E_T\!\!\!\!\!\!/~$ cross-sections in different SRs defined by the ATLAS collaboration \cite{ATLAS-CONF-2019-040} (see section
2 and Table~\ref{SRs}). The {\em nmUED} predictions for the visible cross-sections for the benchmark points (listed in table~\ref{nmUEDBP}) are presented in Table~\ref{SRs}. Fig.~\ref{exc1800} and \ref{exc1900} show the visible cross-sections in signal regions SR2j-1.6 
(top left panel), SR2j-2.2 (top right panel) and SR4j-1.0 (bottom panel) as a function of $R_Q$ and $R_G$ for $R^{-1}~=~$1.8 TeV (Fig. \ref{exc1800}) and 1.9 TeV (Fig. \ref{exc1900}). Clearly, the reddish cells of Fig.~\ref{exc1800} and \ref{exc1900} correspond
to {\em nmUED} cross-sections which are larger than the ATLAS observed 95\% CL upper bound on $\langle \epsilon \sigma\rangle_{obs}^{95}$ (see Table~\ref{SRs}) in the respective signal regions and hence, the corresponding parameter points are ruled out.}

The complementarity of SRs is clearly visible in Fig.~\ref{exc1800}. While, SR2j-2.2 and SR2j-1.6 signal regions are more effective to probe the low (represented by BP$^{nm}_1$) and high (represented by BP$^{nm}_2$) $R_Q$ regions, respectively; the intermediate (represented by BP$^{nm}_3$) $R_Q$ region is susceptible to 
SR4j-1.0 (see Figs.~\ref{exc1800}, \ref{exc1900} and Table~\ref{SRs}). This particular pattern of effectiveness of different SRs in different parts of $R_Q$--$R_G$ plane can be understood from the characteristics of multijet $+$ $E_T\!\!\!\!\!\!/~$ signatures in different parts of $R_Q$--$R_G$ plane.  For example, the region characterized by higher values for both $R_Q$ and $R_G$ (top-right corner of $R_Q$--$R_G$ plane and represented by BP$_2^{nm}$) gives rise to a {\em nmUED} scenario with nearly-degenerate masses for the level-1 quarks, gluons and the LKP (see Table~\ref{nmUEDBP}). Although, the pair and associated production cross-sections of $G^{(1)}$ and $Q^{(1)}$ in this region are large, the final state jets are usually too soft to pass the preselection criteria. The production of $G^{(1)}$ and $Q^{(1)}$ in association with a hard ISR jet gives rise to a monojet $+$ $E_T\!\!\!\!\!\!/~$ signature. It has already been discussed in the context of {\em mUED} phenomenology  that the selection criteria of SR2j-2.2 is essentially a monojet like selection criteria and hence, effective to probe this part of $R_Q$--$R_G$ plane. On the other hand, in the region characterized by low $R_Q$ and/or low $R_G$ (represented by BP$_1^{nm}$ and BP$_3^{nm}$ in Table~\ref{nmUEDBP}), the level-1 quarks and/or gluons are sufficiently heavy compared to the LKP and hence, give rise to hard jets at the LHC. While the pair production of $G^{(1)}$ leads to four hard jets at parton level, two hard jets arise from the pair production of $Q^{(1)}$. Therefore, this part of $R_Q$--$R_G$ plane is susceptible to both SR2j-1.6 and SR4j-1.0. However, our analysis shows that SR2j-1.6 is more efficient to probe this region.

While, all parts of $R_Q$--$R_G$ plane are ruled out from complementary signal regions for $R^{-1}~=~1.8$ TeV (see Fig.~\ref{exc1800}), the ATLAS multijet search can probe only some part of $R_Q$--$R_G$ plane for $R^{-1}~=~1.9$ TeV (see Fig.~\ref{exc1900}). Our final results are summarized in Fig.~\ref{excl_nmued} which shows the lower bounds on $R^{-1}$ in different parts of $R_{Q}$--$R_G$ plane. The signal regions which leads to those lower bounds are also presented in Fig.~\ref{excl_nmued}. For $(R_Q,~R_G)~=~(-0.1,-0.1)$, $R^{-1}$ below 2.15 TeV is ruled out from SR2j-2.2. Whereas, the bound on $R^{-1}$ could be as low as 1.79 TeV for $(R_Q,~R_G)~=~(-0.5,-0.3)$.

\section{Summary and Conclusion}\label{conclusion}
To summarize, we have studied one Universal Extra Dimension scenarios
against the dataset recorded by the ATLAS collaboration
in proton-proton collisions at a center of mass energy $\sqrt{s} = 13$ TeV, corresponding to an integrated luminosity
of 139 fb$^{-1}$. The phenomenology of the minimal version of UED
({\em mUED}) is completely determined by the two new parameters: namely
the compactification radius $R$ and the cut-off scale $\Lambda$.
Our study clearly shows that {\em mUED} parameter space is completely
ruled out by the ATLAS multijet $+~E_T\!\!\!\!\!/~$ analysis together with
the dark matter relic density data.
Next, we bring in boundary-localized terms (with $R_G$ and $R_Q$ as BLT parameters for gluon and quark fields, respectively) as an extension of {\em mUED}, called non-minimal UED. Introduction of such terms
alters the phenomenology substantially. Mass spectrum in {\em nmUED} is
determined by the transcendental equations coming from the boundary
terms. Some interaction vertices are also
altered as a result of integration of extra dimensional mode functions
of the concerned particles. We have discussed strong production
cross sections for the variations of the gluon and quark BLT
parameters. We have performed a detailed
cut-based analysis emulating the ATLAS multijet $+~E_T\!\!\!\!\!/~$
channel. Excluded regions of {\em nmUED} parameter space are shown
in terms of $R^{-1}$, $R_G$ and $R_Q$.


\section{Acknowledgements} TJ thanks Ipsita Saha, Sujoy Shil, Nabanita Ganguly, Sanchari Bhattacharyya and Tisa Biswas for important discussions and many technical helps regarding collider part. TJ also thanks Swagata Ghosh for discussion regarding FeynRules. The authors acknowledge that the simulations and computations were supported in part by the SAMKHYA: High Performance Computing Facility provided by Institute of Physics, Bhubaneswar. KG acknowledges the support from the DST/INSPIRE Research Grant [DST/INSPIRE/04/2014/002158] and SERB Core Research Grant [CRG/2019/006831].
 \bibliographystyle{JHEP}
 \bibliography{nmued.bbl}

\providecommand{\href}[2]{#2}\begingroup\raggedright\begin{thebibliography}{100}

\bibitem{Aad:2012tfa}
{\bf ATLAS} Collaboration, G.~Aad et~al., {\it {Observation of a new particle
  in the search for the Standard Model Higgs boson with the ATLAS detector at
  the LHC}},  {\em Phys. Lett.} {\bf B716} (2012) 1--29,
  [\href{http://arxiv.org/abs/1207.7214}{{\tt arXiv:1207.7214}}].

\bibitem{Chatrchyan:2012xdj}
{\bf CMS} Collaboration, S.~Chatrchyan et~al., {\it {Observation of a new boson
  at a mass of 125 GeV with the CMS experiment at the LHC}},  {\em Phys. Lett.}
  {\bf B716} (2012) 30--61, [\href{http://arxiv.org/abs/1207.7235}{{\tt
  arXiv:1207.7235}}].

\bibitem{ArkaniHamed:1998rs}
N.~Arkani-Hamed, S.~Dimopoulos, and G.~Dvali, {\it {The Hierarchy problem and
  new dimensions at a millimeter}},  {\em Phys. Lett. B} {\bf 429} (1998)
  263--272, [\href{http://arxiv.org/abs/hep-ph/9803315}{{\tt hep-ph/9803315}}].

\bibitem{ArkaniHamed:1998nn}
N.~Arkani-Hamed, S.~Dimopoulos, and G.~Dvali, {\it {Phenomenology, astrophysics
  and cosmology of theories with submillimeter dimensions and TeV scale quantum
  gravity}},  {\em Phys. Rev. D} {\bf 59} (1999) 086004,
  [\href{http://arxiv.org/abs/hep-ph/9807344}{{\tt hep-ph/9807344}}].

\bibitem{Randall:1999ee}
L.~Randall and R.~Sundrum, {\it {A Large mass hierarchy from a small extra
  dimension}},  {\em Phys. Rev. Lett.} {\bf 83} (1999) 3370--3373,
  [\href{http://arxiv.org/abs/hep-ph/9905221}{{\tt hep-ph/9905221}}].

\bibitem{Randall:1999vf}
L.~Randall and R.~Sundrum, {\it {An Alternative to compactification}},  {\em
  Phys. Rev. Lett.} {\bf 83} (1999) 4690--4693,
  [\href{http://arxiv.org/abs/hep-th/9906064}{{\tt hep-th/9906064}}].

\bibitem{Dienes:1998sb}
K.~R. Dienes, E.~Dudas, and T.~Gherghetta, {\it {Neutrino oscillations without
  neutrino masses or heavy mass scales: A Higher dimensional seesaw
  mechanism}},  {\em Nucl. Phys. B} {\bf 557} (1999) 25,
  [\href{http://arxiv.org/abs/hep-ph/9811428}{{\tt hep-ph/9811428}}].

\bibitem{Dienes:1998vh}
K.~R. Dienes, E.~Dudas, and T.~Gherghetta, {\it {Extra space-time dimensions
  and unification}},  {\em Phys. Lett. B} {\bf 436} (1998) 55--65,
  [\href{http://arxiv.org/abs/hep-ph/9803466}{{\tt hep-ph/9803466}}].

\bibitem{ArkaniHamed:1999dc}
N.~Arkani-Hamed and M.~Schmaltz, {\it {Hierarchies without symmetries from
  extra dimensions}},  {\em Phys. Rev. D} {\bf 61} (2000) 033005,
  [\href{http://arxiv.org/abs/hep-ph/9903417}{{\tt hep-ph/9903417}}].

\bibitem{ANTONIADIS1990377}
I.~Antoniadis, {\it A possible new dimension at a few tev},  {\em Physics
  Letters B} {\bf 246} (1990), no.~3 377 -- 384.

\bibitem{PhysRevD.64.035002}
T.~Appelquist, H.-C. Cheng, and B.~A. Dobrescu, {\it Bounds on universal extra
  dimensions},  {\em Phys. Rev. D} {\bf 64} (Jun, 2001) 035002.

\bibitem{PhysRevD.66.056006}
H.-C. Cheng, K.~T. Matchev, and M.~Schmaltz, {\it Bosonic supersymmetry?
  getting fooled at the cern lhc},  {\em Phys. Rev. D} {\bf 66} (Sep, 2002)
  056006.

\bibitem{ArkaniHamed:2000hv}
N.~Arkani-Hamed, H.-C. Cheng, B.~A. Dobrescu, and L.~J. Hall, {\it
  {Selfbreaking of the standard model gauge symmetry}},  {\em Phys. Rev. D}
  {\bf 62} (2000) 096006, [\href{http://arxiv.org/abs/hep-ph/0006238}{{\tt
  hep-ph/0006238}}].

\bibitem{Servant:2002aq}
G.~Servant and T.~M. Tait, {\it {Is the lightest Kaluza-Klein particle a viable
  dark matter candidate?}},  {\em Nucl. Phys. B} {\bf 650} (2003) 391--419,
  [\href{http://arxiv.org/abs/hep-ph/0206071}{{\tt hep-ph/0206071}}].

\bibitem{Kakizaki:2006dz}
M.~Kakizaki, S.~Matsumoto, and M.~Senami, {\it {Relic abundance of dark matter
  in the minimal universal extra dimension model}},  {\em Phys. Rev. D} {\bf
  74} (2006) 023504, [\href{http://arxiv.org/abs/hep-ph/0605280}{{\tt
  hep-ph/0605280}}].

\bibitem{Dienes:1998vg}
K.~R. Dienes, E.~Dudas, and T.~Gherghetta, {\it {Grand unification at
  intermediate mass scales through extra dimensions}},  {\em Nucl. Phys. B}
  {\bf 537} (1999) 47--108, [\href{http://arxiv.org/abs/hep-ph/9806292}{{\tt
  hep-ph/9806292}}].

\bibitem{Appelquist:2001mj}
T.~Appelquist, B.~A. Dobrescu, E.~Ponton, and H.-U. Yee, {\it {Proton stability
  in six-dimensions}},  {\em Phys. Rev. Lett.} {\bf 87} (2001) 181802,
  [\href{http://arxiv.org/abs/hep-ph/0107056}{{\tt hep-ph/0107056}}].

\bibitem{Dobrescu:2001ae}
B.~A. Dobrescu and E.~Poppitz, {\it {Number of fermion generations derived from
  anomaly cancellation}},  {\em Phys. Rev. Lett.} {\bf 87} (2001) 031801,
  [\href{http://arxiv.org/abs/hep-ph/0102010}{{\tt hep-ph/0102010}}].

\bibitem{Appelquist:2000nn}
T.~Appelquist, H.-C. Cheng, and B.~A. Dobrescu, {\it {Bounds on universal extra
  dimensions}},  {\em Phys. Rev. D} {\bf 64} (2001) 035002,
  [\href{http://arxiv.org/abs/hep-ph/0012100}{{\tt hep-ph/0012100}}].

\bibitem{Belyaev:2012ai}
A.~Belyaev, M.~Brown, J.~Moreno, and C.~Papineau, {\it {Discovering Minimal
  Universal Extra Dimensions (MUED) at the LHC}},  {\em JHEP} {\bf 06} (2013)
  080, [\href{http://arxiv.org/abs/1212.4858}{{\tt arXiv:1212.4858}}].

\bibitem{Kakuda:2013kba}
T.~Kakuda, K.~Nishiwaki, K.-y. Oda, and R.~Watanabe, {\it {Universal extra
  dimensions after Higgs discovery}},  {\em Phys. Rev. D} {\bf 88} (2013)
  035007, [\href{http://arxiv.org/abs/1305.1686}{{\tt arXiv:1305.1686}}].

\bibitem{Belanger:2012mc}
G.~Belanger, A.~Belyaev, M.~Brown, M.~Kakizaki, and A.~Pukhov, {\it {Testing
  Minimal Universal Extra Dimensions Using Higgs Boson Searches at the LHC}},
  {\em Phys. Rev. D} {\bf 87} (2013), no.~1 016008,
  [\href{http://arxiv.org/abs/1207.0798}{{\tt arXiv:1207.0798}}].

\bibitem{Ghosh:2018mck}
K.~Ghosh, D.~Karabacak, and S.~Nandi, {\it {Universal Extra Dimension models
  with gravity mediated decays after LHC Run II data}},  {\em Phys. Lett. B}
  {\bf 788} (2019) 388--395, [\href{http://arxiv.org/abs/1805.11124}{{\tt
  arXiv:1805.11124}}].

\bibitem{Ghosh:2008dp}
K.~Ghosh and A.~Datta, {\it {Probing two Universal Extra Dimensions at
  International Linear Collider}},  {\em Phys. Lett. B} {\bf 665} (2008)
  369--373, [\href{http://arxiv.org/abs/0802.2162}{{\tt arXiv:0802.2162}}].

\bibitem{Ghosh:2008ix}
K.~Ghosh and A.~Datta, {\it {Phenomenology of spinless adjoints in two
  Universal Extra Dimensions}},  {\em Nucl. Phys. B} {\bf 800} (2008) 109--126,
  [\href{http://arxiv.org/abs/0801.0943}{{\tt arXiv:0801.0943}}].

\bibitem{Beuria:2017jez}
J.~Beuria, A.~Datta, D.~Debnath, and K.~T. Matchev, {\it {LHC Collider
  Phenomenology of Minimal Universal Extra Dimensions}},  {\em Comput. Phys.
  Commun.} {\bf 226} (2018) 187--205,
  [\href{http://arxiv.org/abs/1702.00413}{{\tt arXiv:1702.00413}}].

\bibitem{Dey:2014ana}
U.~K. Dey and A.~Raychaudhuri, {\it {KK-number non-conserving decays: Signal of
  $n =$ 2 excitations of extra-dimensional models at the LHC}},  {\em Nucl.
  Phys. B} {\bf 893} (2015) 408--419,
  [\href{http://arxiv.org/abs/1410.1463}{{\tt arXiv:1410.1463}}].

\bibitem{Edelhauser:2013lia}
L.~Edelh\"auser, T.~Flacke, and M.~Kr\"amer, {\it {Constraints on models with
  universal extra dimensions from dilepton searches at the LHC}},  {\em JHEP}
  {\bf 08} (2013) 091, [\href{http://arxiv.org/abs/1302.6076}{{\tt
  arXiv:1302.6076}}].

\bibitem{Flacke:2012ke}
T.~Flacke, A.~Menon, and Z.~Sullivan, {\it {Constraints on UED from W'
  searches}},  {\em Phys. Rev. D} {\bf 86} (2012) 093006,
  [\href{http://arxiv.org/abs/1207.4472}{{\tt arXiv:1207.4472}}].

\bibitem{Huang:2012kz}
G.-Y. Huang, K.~Kong, and S.~C. Park, {\it {Bounds on the Fermion-Bulk Masses
  in Models with Universal Extra Dimensions}},  {\em JHEP} {\bf 06} (2012) 099,
  [\href{http://arxiv.org/abs/1204.0522}{{\tt arXiv:1204.0522}}].

\bibitem{Flacke:2011nb}
T.~Flacke and C.~Pasold, {\it {Constraints on split-UED from Electroweak
  Precision Tests}},  {\em Phys. Rev. D} {\bf 85} (2012) 126007,
  [\href{http://arxiv.org/abs/1111.7250}{{\tt arXiv:1111.7250}}].

\bibitem{Nishiwaki:2011gm}
K.~Nishiwaki, K.-y. Oda, N.~Okuda, and R.~Watanabe, {\it {Heavy Higgs at
  Tevatron and LHC in Universal Extra Dimension Models}},  {\em Phys. Rev. D}
  {\bf 85} (2012) 035026, [\href{http://arxiv.org/abs/1108.1765}{{\tt
  arXiv:1108.1765}}].

\bibitem{Murayama:2011hj}
H.~Murayama, M.~M. Nojiri, and K.~Tobioka, {\it {Improved discovery of a nearly
  degenerate model: MUED using MT2 at the LHC}},  {\em Phys. Rev. D} {\bf 84}
  (2011) 094015, [\href{http://arxiv.org/abs/1107.3369}{{\tt
  arXiv:1107.3369}}].

\bibitem{Choudhury:2011jk}
D.~Choudhury, A.~Datta, D.~K. Ghosh, and K.~Ghosh, {\it {Exploring two
  Universal Extra Dimensions at the CERN LHC}},  {\em JHEP} {\bf 04} (2012)
  057, [\href{http://arxiv.org/abs/1109.1400}{{\tt arXiv:1109.1400}}].

\bibitem{Ghosh:2010tp}
K.~Ghosh, S.~Mukhopadhyay, and B.~Mukhopadhyaya, {\it {Discrimination of low
  missing energy look-alikes at the LHC}},  {\em JHEP} {\bf 10} (2010) 096,
  [\href{http://arxiv.org/abs/1007.4012}{{\tt arXiv:1007.4012}}].

\bibitem{Bhattacherjee:2010vm}
B.~Bhattacherjee and K.~Ghosh, {\it {Search for the minimal universal extra
  dimension model at the LHC with $\sqrt{s}$=7 TeV}},  {\em Phys. Rev. D} {\bf
  83} (2011) 034003, [\href{http://arxiv.org/abs/1006.3043}{{\tt
  arXiv:1006.3043}}].

\bibitem{Bertone:2010ww}
G.~Bertone, K.~Kong, R.~Ruiz~de Austri, and R.~Trotta, {\it {Global fits of the
  Minimal Universal Extra Dimensions scenario}},  {\em Phys. Rev. D} {\bf 83}
  (2011) 036008, [\href{http://arxiv.org/abs/1010.2023}{{\tt
  arXiv:1010.2023}}].

\bibitem{Freitas:2007rh}
A.~Freitas and K.~Kong, {\it {Two universal extra dimensions and spinless
  photons at the ILC}},  {\em JHEP} {\bf 02} (2008) 068,
  [\href{http://arxiv.org/abs/0711.4124}{{\tt arXiv:0711.4124}}].

\bibitem{Dobrescu:2007ec}
B.~A. Dobrescu, D.~Hooper, K.~Kong, and R.~Mahbubani, {\it {Spinless photon
  dark matter from two universal extra dimensions}},  {\em JCAP} {\bf 10}
  (2007) 012, [\href{http://arxiv.org/abs/0706.3409}{{\tt arXiv:0706.3409}}].

\bibitem{Macesanu:2002db}
C.~Macesanu, C.~McMullen, and S.~Nandi, {\it {Collider implications of
  universal extra dimensions}},  {\em Phys. Rev. D} {\bf 66} (2002) 015009,
  [\href{http://arxiv.org/abs/hep-ph/0201300}{{\tt hep-ph/0201300}}].

\bibitem{Ghosh:2012zc}
K.~Ghosh and K.~Huitu, {\it {Constraints on Universal Extra Dimension models
  with gravity mediated decays from ATLAS diphoton search}},  {\em JHEP} {\bf
  06} (2012) 042, [\href{http://arxiv.org/abs/1203.1551}{{\tt
  arXiv:1203.1551}}].

\bibitem{Choudhury:2009kz}
D.~Choudhury, A.~Datta, and K.~Ghosh, {\it {Deciphering Universal Extra
  Dimension from the top quark signals at the CERN LHC}},  {\em JHEP} {\bf 08}
  (2010) 051, [\href{http://arxiv.org/abs/0911.4064}{{\tt arXiv:0911.4064}}].

\bibitem{Rizzo:2001sd}
T.~G. Rizzo, {\it {Probes of universal extra dimensions at colliders}},  {\em
  Phys. Rev. D} {\bf 64} (2001) 095010,
  [\href{http://arxiv.org/abs/hep-ph/0106336}{{\tt hep-ph/0106336}}].

\bibitem{Muck:2003kx}
A.~Muck, A.~Pilaftsis, and R.~Ruckl, {\it {Probing minimal 5-D extensions of
  the standard model: From LEP to an e+ e- linear collider}},  {\em Nucl. Phys.
  B} {\bf 687} (2004) 55--75, [\href{http://arxiv.org/abs/hep-ph/0312186}{{\tt
  hep-ph/0312186}}].

\bibitem{Bhattacharyya:2005vm}
G.~Bhattacharyya, P.~Dey, A.~Kundu, and A.~Raychaudhuri, {\it {Probing
  universal extra dimension at the international linear collider}},  {\em Phys.
  Lett. B} {\bf 628} (2005) 141--147,
  [\href{http://arxiv.org/abs/hep-ph/0502031}{{\tt hep-ph/0502031}}].

\bibitem{Battaglia:2005zf}
M.~Battaglia, A.~Datta, A.~De~Roeck, K.~Kong, and K.~T. Matchev, {\it
  {Contrasting supersymmetry and universal extra dimensions at the clic
  multi-TeV e+ e- collider}},  {\em JHEP} {\bf 07} (2005) 033,
  [\href{http://arxiv.org/abs/hep-ph/0502041}{{\tt hep-ph/0502041}}].

\bibitem{Bhattacherjee:2005qe}
B.~Bhattacherjee and A.~Kundu, {\it {The International linear collider as a
  Kaluza-Klein factory}},  {\em Phys. Lett. B} {\bf 627} (2005) 137--144,
  [\href{http://arxiv.org/abs/hep-ph/0508170}{{\tt hep-ph/0508170}}].

\bibitem{Datta:2005zs}
A.~Datta, K.~Kong, and K.~T. Matchev, {\it {Discrimination of supersymmetry and
  universal extra dimensions at hadron colliders}},  {\em Phys. Rev. D} {\bf
  72} (2005) 096006, [\href{http://arxiv.org/abs/hep-ph/0509246}{{\tt
  hep-ph/0509246}}]. [Erratum: Phys.Rev.D 72, 119901 (2005)].

\bibitem{Datta:2005vx}
A.~Datta, G.~L. Kane, and M.~Toharia, {\it {Is it SUSY?}},
  \href{http://arxiv.org/abs/hep-ph/0510204}{{\tt hep-ph/0510204}}.

\bibitem{Kong:2010qd}
K.~Kong, S.~C. Park, and T.~G. Rizzo, {\it {A vector-like fourth generation
  with a discrete symmetry from Split-UED}},  {\em JHEP} {\bf 07} (2010) 059,
  [\href{http://arxiv.org/abs/1004.4635}{{\tt arXiv:1004.4635}}].

\bibitem{Chen:2009gz}
C.-R. Chen, M.~M. Nojiri, S.~C. Park, J.~Shu, and M.~Takeuchi, {\it {Dark
  matter and collider phenomenology of split-UED}},  {\em JHEP} {\bf 09} (2009)
  078, [\href{http://arxiv.org/abs/0903.1971}{{\tt arXiv:0903.1971}}].

\bibitem{Park:2009cs}
S.~C. Park and J.~Shu, {\it {Split Universal Extra Dimensions and Dark
  Matter}},  {\em Phys. Rev. D} {\bf 79} (2009) 091702,
  [\href{http://arxiv.org/abs/0901.0720}{{\tt arXiv:0901.0720}}].

\bibitem{delAguila:2003bh}
F.~del Aguila, M.~Perez-Victoria, and J.~Santiago, {\it {Bulk fields with
  general brane kinetic terms}},  {\em JHEP} {\bf 02} (2003) 051,
  [\href{http://arxiv.org/abs/hep-th/0302023}{{\tt hep-th/0302023}}].

\bibitem{Carena:2002me}
M.~Carena, T.~M. Tait, and C.~Wagner, {\it {Branes and Orbifolds are Opaque}},
  {\em Acta Phys. Polon. B} {\bf 33} (2002) 2355,
  [\href{http://arxiv.org/abs/hep-ph/0207056}{{\tt hep-ph/0207056}}].

\bibitem{Cheng:2002iz}
H.-C. Cheng, K.~T. Matchev, and M.~Schmaltz, {\it {Radiative corrections to
  Kaluza-Klein masses}},  {\em Phys. Rev. D} {\bf 66} (2002) 036005,
  [\href{http://arxiv.org/abs/hep-ph/0204342}{{\tt hep-ph/0204342}}].

\bibitem{Choudhury:2016tff}
D.~Choudhury and K.~Ghosh, {\it {Bounds on Universal Extra Dimension from LHC
  Run I and II data}},  {\em Phys. Lett. B} {\bf 763} (2016) 155--160,
  [\href{http://arxiv.org/abs/1606.04084}{{\tt arXiv:1606.04084}}].

\bibitem{Deutschmann:2017bth}
N.~Deutschmann, T.~Flacke, and J.~S. Kim, {\it {Current LHC Constraints on
  Minimal Universal Extra Dimensions}},  {\em Phys. Lett. B} {\bf 771} (2017)
  515--520, [\href{http://arxiv.org/abs/1702.00410}{{\tt arXiv:1702.00410}}].

\bibitem{Cornell:2014jza}
J.~M. Cornell, S.~Profumo, and W.~Shepherd, {\it {Dark matter in minimal
  universal extra dimensions with a stable vacuum and the
  \textquotedblleft{}right\textquotedblright{} Higgs boson}},  {\em Phys. Rev.
  D} {\bf 89} (2014), no.~5 056005, [\href{http://arxiv.org/abs/1401.7050}{{\tt
  arXiv:1401.7050}}].

\bibitem{Komatsu:2010fb}
{\bf WMAP} Collaboration, E.~Komatsu et~al., {\it {Seven-Year Wilkinson
  Microwave Anisotropy Probe (WMAP) Observations: Cosmological
  Interpretation}},  {\em Astrophys. J. Suppl.} {\bf 192} (2011) 18,
  [\href{http://arxiv.org/abs/1001.4538}{{\tt arXiv:1001.4538}}].

\bibitem{Ade:2015xua}
{\bf Planck} Collaboration, P.~Ade et~al., {\it {Planck 2015 results. XIII.
  Cosmological parameters}},  {\em Astron. Astrophys.} {\bf 594} (2016) A13,
  [\href{http://arxiv.org/abs/1502.01589}{{\tt arXiv:1502.01589}}].

\bibitem{Flacke:2008ne}
T.~Flacke, A.~Menon, and D.~J. Phalen, {\it {Non-minimal universal extra
  dimensions}},  {\em Phys. Rev. D} {\bf 79} (2009) 056009,
  [\href{http://arxiv.org/abs/0811.1598}{{\tt arXiv:0811.1598}}].

\bibitem{Datta:2012tv}
A.~Datta, K.~Nishiwaki, and S.~Niyogi, {\it {Non-minimal Universal Extra
  Dimensions: The Strongly Interacting Sector at the Large Hadron Collider}},
  {\em JHEP} {\bf 11} (2012) 154, [\href{http://arxiv.org/abs/1206.3987}{{\tt
  arXiv:1206.3987}}].

\bibitem{Flacke:2013nta}
T.~Flacke, K.~Kong, and S.~C. Park, {\it {126 GeV Higgs in Next-to-Minimal
  Universal Extra Dimensions}},  {\em Phys. Lett. B} {\bf 728} (2014) 262--267,
  [\href{http://arxiv.org/abs/1309.7077}{{\tt arXiv:1309.7077}}].

\bibitem{Datta:2013yaa}
A.~Datta, K.~Nishiwaki, and S.~Niyogi, {\it {Non-minimal Universal Extra
  Dimensions with Brane Local Terms: The Top Quark Sector}},  {\em JHEP} {\bf
  01} (2014) 104, [\href{http://arxiv.org/abs/1310.6994}{{\tt
  arXiv:1310.6994}}].

\bibitem{Flacke:2013pla}
T.~Flacke, K.~Kong, and S.~C. Park, {\it {Phenomenology of Universal Extra
  Dimensions with Bulk-Masses and Brane-Localized Terms}},  {\em JHEP} {\bf 05}
  (2013) 111, [\href{http://arxiv.org/abs/1303.0872}{{\tt arXiv:1303.0872}}].

\bibitem{Shaw:2017whr}
A.~Shaw, {\it {Status of exclusion limits of the KK-parity non-conserving
  resonance production with updated 13 TeV LHC}},  {\em Acta Phys. Polon. B}
  {\bf 49} (2018) 1421, [\href{http://arxiv.org/abs/1709.08077}{{\tt
  arXiv:1709.08077}}].

\bibitem{Datta:2013ufa}
A.~Datta, U.~K. Dey, A.~Raychaudhuri, and A.~Shaw, {\it {Universal
  Extra-Dimensional models with boundary terms: Probing at the LHC}},  {\em
  Nucl. Phys. B Proc. Suppl.} {\bf 251-252} (2014) 39--44,
  [\href{http://arxiv.org/abs/1312.2312}{{\tt arXiv:1312.2312}}].

\bibitem{Dey:2013cqa}
U.~K. Dey and T.~S. Ray, {\it {Constraining minimal and nonminimal universal
  extra dimension models with Higgs couplings}},  {\em Phys. Rev. D} {\bf 88}
  (2013), no.~5 056016, [\href{http://arxiv.org/abs/1305.1016}{{\tt
  arXiv:1305.1016}}].

\bibitem{Ghosh:2014uwa}
K.~Ghosh, D.~Karabacak, and S.~Nandi, {\it {Constraining Bosonic Supersymmetry
  from Higgs results and 8 TeV ATLAS multi-jets plus missing energy data}},
  {\em JHEP} {\bf 09} (2014) 076, [\href{http://arxiv.org/abs/1402.5939}{{\tt
  arXiv:1402.5939}}].

\bibitem{Datta:2013nua}
A.~Datta, U.~K. Dey, A.~Raychaudhuri, and A.~Shaw, {\it {Boundary Localized
  Terms in Universal Extra-Dimensional Models through a Dark Matter
  perspective}},  {\em Phys. Rev. D} {\bf 88} (2013) 016011,
  [\href{http://arxiv.org/abs/1305.4507}{{\tt arXiv:1305.4507}}].

\bibitem{Flacke:2017xsv}
T.~Flacke, D.~W. Kang, K.~Kong, G.~Mohlabeng, and S.~C. Park, {\it {Electroweak
  Kaluza-Klein Dark Matter}},  {\em JHEP} {\bf 04} (2017) 041,
  [\href{http://arxiv.org/abs/1702.02949}{{\tt arXiv:1702.02949}}].

\bibitem{Datta:2015aka}
A.~Datta and A.~Shaw, {\it {Nonminimal universal extra dimensional model
  confronts B$_s\to \mu^+\mu^-$}},  {\em Phys. Rev. D} {\bf 93} (2016), no.~5
  055048, [\href{http://arxiv.org/abs/1506.08024}{{\tt arXiv:1506.08024}}].

\bibitem{Datta:2016flx}
A.~Datta and A.~Shaw, {\it {Effects of non-minimal Universal Extra Dimension on
  $B\rightarrow X_s\gamma$}},  {\em Phys. Rev. D} {\bf 95} (2017), no.~1
  015033, [\href{http://arxiv.org/abs/1610.09924}{{\tt arXiv:1610.09924}}].

\bibitem{Biswas:2017vhc}
A.~Biswas, A.~Shaw, and S.~K. Patra, {\it {$\mathcal{R}(D^{(*)})$ anomalies in
  light of a nonminimal universal extra dimension}},  {\em Phys. Rev. D} {\bf
  97} (2018), no.~3 035019, [\href{http://arxiv.org/abs/1708.08938}{{\tt
  arXiv:1708.08938}}].

\bibitem{Jha:2016sre}
T.~Jha, {\it {Unitarity Constraints on non-minimal Universal Extra Dimensional
  Model}},  {\em J. Phys. G} {\bf 45} (2018), no.~11 115002,
  [\href{http://arxiv.org/abs/1604.02481}{{\tt arXiv:1604.02481}}].

\bibitem{Jha:2014faa}
T.~Jha and A.~Datta, {\it {$ Z\to b\overline{b} $ in non-minimal Universal
  Extra Dimensional Model}},  {\em JHEP} {\bf 03} (2015) 012,
  [\href{http://arxiv.org/abs/1410.5098}{{\tt arXiv:1410.5098}}].

\bibitem{Jha:2017oek}
T.~Jha, {\em {Exploration in extra dimension in the era of LHC and beyond}}.
\newblock PhD thesis, Calcutta U., 2017.

\bibitem{Dasgupta:2018nzt}
S.~Dasgupta, U.~K. Dey, T.~Jha, and T.~S. Ray, {\it {Status of a flavor-maximal
  nonminimal universal extra dimension model}},  {\em Phys. Rev. D} {\bf 98}
  (2018), no.~5 055006, [\href{http://arxiv.org/abs/1801.09722}{{\tt
  arXiv:1801.09722}}].

\bibitem{Ganguly:2018pzs}
N.~Ganguly and A.~Datta, {\it {Exploring non minimal Universal Extra
  Dimensional Model at the LHC}},  {\em JHEP} {\bf 10} (2018) 072,
  [\href{http://arxiv.org/abs/1808.08801}{{\tt arXiv:1808.08801}}].

\bibitem{Dey:2016cve}
U.~K. Dey and T.~Jha, {\it {Rare top decays in minimal and nonminimal universal
  extra dimension models}},  {\em Phys. Rev. D} {\bf 94} (2016), no.~5 056011,
  [\href{http://arxiv.org/abs/1602.03286}{{\tt arXiv:1602.03286}}].

\bibitem{Chiang:2018oyd}
C.-W. Chiang, U.~K. Dey, and T.~Jha, {\it {$t \rightarrow cg$ and $t
  \rightarrow cZ$ in universal extra-dimensional models}},  {\em Eur. Phys. J.
  Plus} {\bf 134} (2019), no.~5 210,
  [\href{http://arxiv.org/abs/1807.01481}{{\tt arXiv:1807.01481}}].

\bibitem{Datta:2014yua}
A.~Datta, U.~K. Dey, A.~Raychaudhuri, and A.~Shaw, {\it {Universal extra
  dimensions : life with BLKTs}},  {\em J. Phys. Conf. Ser.} {\bf 481} (2014)
  012006.

\bibitem{ATLAS-CONF-2019-040}
{\bf ATLAS Collaboration} Collaboration, {\it {Search for squarks and gluinos
  in final states with jets and missing transverse momentum using 139 fb$^{-1}$
  of $\sqrt{s}$ =13 TeV $pp$ collision data with the ATLAS detector}},  Tech.
  Rep. ATLAS-CONF-2019-040, CERN, Geneva, Aug, 2019.

\bibitem{Cacciari:2008gp}
M.~Cacciari, G.~P. Salam, and G.~Soyez, {\it {The Anti-k(t) jet clustering
  algorithm}},  {\em JHEP} {\bf 04} (2008) 063,
  [\href{http://arxiv.org/abs/0802.1189}{{\tt arXiv:0802.1189}}].

\bibitem{Cacciari:2011ma}
M.~Cacciari, G.~P. Salam, and G.~Soyez, {\it {FastJet User Manual}},  {\em Eur.
  Phys. J.} {\bf C72} (2012) 1896, [\href{http://arxiv.org/abs/1111.6097}{{\tt
  arXiv:1111.6097}}].

\bibitem{Bjorken:1969wi}
J.~Bjorken and S.~J. Brodsky, {\it {Statistical Model for electron-Positron
  Annihilation Into Hadrons}},  {\em Phys. Rev. D} {\bf 1} (1970) 1416--1420.

\bibitem{Alwall:2014hca}
J.~Alwall, R.~Frederix, S.~Frixione, V.~Hirschi, F.~Maltoni, O.~Mattelaer,
  H.~S. Shao, T.~Stelzer, P.~Torrielli, and M.~Zaro, {\it {The automated
  computation of tree-level and next-to-leading order differential cross
  sections, and their matching to parton shower simulations}},  {\em JHEP} {\bf
  07} (2014) 079, [\href{http://arxiv.org/abs/1405.0301}{{\tt
  arXiv:1405.0301}}].

\bibitem{Ball:2014uwa}
{\bf NNPDF} Collaboration, R.~D. Ball et~al., {\it {Parton distributions for
  the LHC Run II}},  {\em JHEP} {\bf 04} (2015) 040,
  [\href{http://arxiv.org/abs/1410.8849}{{\tt arXiv:1410.8849}}].

\bibitem{Sjostrand:2007gs}
T.~Sjostrand, S.~Mrenna, and P.~Z. Skands, {\it {A Brief Introduction to PYTHIA
  8.1}},  {\em Comput. Phys. Commun.} {\bf 178} (2008) 852--867,
  [\href{http://arxiv.org/abs/0710.3820}{{\tt arXiv:0710.3820}}].

\bibitem{Sjostrand:2014zea}
T.~Sjöstrand, S.~Ask, J.~R. Christiansen, R.~Corke, N.~Desai, P.~Ilten,
  S.~Mrenna, S.~Prestel, C.~O. Rasmussen, and P.~Z. Skands, {\it {An
  Introduction to PYTHIA 8.2}},  {\em Comput. Phys. Commun.} {\bf 191} (2015)
  159--177, [\href{http://arxiv.org/abs/1410.3012}{{\tt arXiv:1410.3012}}].

\bibitem{Catani_2001}
S.~Catani, F.~Krauss, B.~R. Webber, and R.~Kuhn, {\it Qcd matrix elements +
  parton showers},  {\em Journal of High Energy Physics} {\bf 2001} (Nov, 2001)
  063–063.

\bibitem{deFavereau:2013fsa}
{\bf DELPHES 3} Collaboration, J.~de~Favereau, C.~Delaere, P.~Demin,
  A.~Giammanco, V.~Lema\^\i{}tre, A.~Mertens, and M.~Selvaggi, {\it {DELPHES 3,
  A modular framework for fast simulation of a generic collider experiment}},
  {\em JHEP} {\bf 02} (2014) 057, [\href{http://arxiv.org/abs/1307.6346}{{\tt
  arXiv:1307.6346}}].

\bibitem{Hooper:2007qk}
D.~Hooper and S.~Profumo, {\it {Dark Matter and Collider Phenomenology of
  Universal Extra Dimensions}},  {\em Phys. Rept.} {\bf 453} (2007) 29--115,
  [\href{http://arxiv.org/abs/hep-ph/0701197}{{\tt hep-ph/0701197}}].

\bibitem{Datta:2012db}
A.~Datta and S.~Raychaudhuri, {\it {Vacuum Stability Constraints and LHC
  Searches for a Model with a Universal Extra Dimension}},  {\em Phys. Rev. D}
  {\bf 87} (2013), no.~3 035018, [\href{http://arxiv.org/abs/1207.0476}{{\tt
  arXiv:1207.0476}}].

\bibitem{Nath:1999aa}
P.~Nath and M.~Yamaguchi, {\it {Effects of Kaluza-Klein excitations on
  (g(mu)-2)}},  {\em Phys. Rev. D} {\bf 60} (1999) 116006,
  [\href{http://arxiv.org/abs/hep-ph/9903298}{{\tt hep-ph/9903298}}].

\bibitem{Agashe:2001ra}
K.~Agashe, N.~Deshpande, and G.~Wu, {\it {Can extra dimensions accessible to
  the SM explain the recent measurement of anomalous magnetic moment of the
  muon?}},  {\em Phys. Lett. B} {\bf 511} (2001) 85--91,
  [\href{http://arxiv.org/abs/hep-ph/0103235}{{\tt hep-ph/0103235}}].

\bibitem{Chakraverty:2002qk}
D.~Chakraverty, K.~Huitu, and A.~Kundu, {\it {Effects of universal extra
  dimensions on B0 - antiB0 mixing}},  {\em Phys. Lett. B} {\bf 558} (2003)
  173--181, [\href{http://arxiv.org/abs/hep-ph/0212047}{{\tt hep-ph/0212047}}].

\bibitem{Buras:2003mk}
A.~J. Buras, A.~Poschenrieder, M.~Spranger, and A.~Weiler, {\it {The Impact of
  universal extra dimensions on B ---\ensuremath{>} X(s) gamma, B
  ---\ensuremath{>} X(s) gluon, B ---\ensuremath{>} X(s) mu+ mu-, K(L)
  ---\ensuremath{>} pi0 e+ e- and epsilon-prime / epsilon}},  {\em Nucl. Phys.
  B} {\bf 678} (2004) 455--490,
  [\href{http://arxiv.org/abs/hep-ph/0306158}{{\tt hep-ph/0306158}}].

\bibitem{Agashe:2001xt}
K.~Agashe, N.~Deshpande, and G.~Wu, {\it {Universal extra dimensions and $b\to
  s \gamma$}},  {\em Phys. Lett. B} {\bf 514} (2001) 309--314,
  [\href{http://arxiv.org/abs/hep-ph/0105084}{{\tt hep-ph/0105084}}].

\bibitem{Oliver:2002up}
J.~Oliver, J.~Papavassiliou, and A.~Santamaria, {\it {Universal extra
  dimensions and Z ---\ensuremath{>} b anti-b}},  {\em Phys. Rev. D} {\bf 67}
  (2003) 056002, [\href{http://arxiv.org/abs/hep-ph/0212391}{{\tt
  hep-ph/0212391}}].

\bibitem{Appelquist:2002wb}
T.~Appelquist and H.-U. Yee, {\it {Universal extra dimensions and the Higgs
  boson mass}},  {\em Phys. Rev. D} {\bf 67} (2003) 055002,
  [\href{http://arxiv.org/abs/hep-ph/0211023}{{\tt hep-ph/0211023}}].

\bibitem{Haisch:2007vb}
U.~Haisch and A.~Weiler, {\it {Bound on minimal universal extra dimensions from
  anti-B ---\ensuremath{>} X(s)gamma}},  {\em Phys. Rev. D} {\bf 76} (2007)
  034014, [\href{http://arxiv.org/abs/hep-ph/0703064}{{\tt hep-ph/0703064}}].

\bibitem{Rizzo:1999br}
T.~G. Rizzo and J.~D. Wells, {\it {Electroweak precision measurements and
  collider probes of the standard model with large extra dimensions}},  {\em
  Phys. Rev. D} {\bf 61} (2000) 016007,
  [\href{http://arxiv.org/abs/hep-ph/9906234}{{\tt hep-ph/9906234}}].

\bibitem{Strumia:1999jm}
A.~Strumia, {\it {Bounds on Kaluza-Klein excitations of the SM vector bosons
  from electroweak tests}},  {\em Phys. Lett. B} {\bf 466} (1999) 107--114,
  [\href{http://arxiv.org/abs/hep-ph/9906266}{{\tt hep-ph/9906266}}].

\bibitem{Carone:1999nz}
C.~D. Carone, {\it {Electroweak constraints on extended models with extra
  dimensions}},  {\em Phys. Rev. D} {\bf 61} (2000) 015008,
  [\href{http://arxiv.org/abs/hep-ph/9907362}{{\tt hep-ph/9907362}}].

\bibitem{Gogoladze:2006br}
I.~Gogoladze and C.~Macesanu, {\it {Precision electroweak constraints on
  Universal Extra Dimensions revisited}},  {\em Phys. Rev. D} {\bf 74} (2006)
  093012, [\href{http://arxiv.org/abs/hep-ph/0605207}{{\tt hep-ph/0605207}}].

\bibitem{Dey:2004gb}
P.~Dey and G.~Bhattacharyya, {\it {A Comparison of ultraviolet sensitivities in
  universal, nonuniversal, and split extra dimensional models}},  {\em Phys.
  Rev. D} {\bf 70} (2004) 116012,
  [\href{http://arxiv.org/abs/hep-ph/0407314}{{\tt hep-ph/0407314}}].

\bibitem{Dey:2003yh}
P.~Dey and G.~Bhattacharyya, {\it {Ultraviolet sensitivity of rare decays in
  nonuniversal extra- dimensional models}},  {\em Phys. Rev. D} {\bf 69} (2004)
  076009, [\href{http://arxiv.org/abs/hep-ph/0309110}{{\tt hep-ph/0309110}}].

\bibitem{Aad:2015zva}
{\bf ATLAS} Collaboration, G.~Aad et~al., {\it {Search for new phenomena in
  final states with an energetic jet and large missing transverse momentum in
  pp collisions at $\sqrt{s}=$8 TeV with the ATLAS detector}},  {\em Eur. Phys.
  J. C} {\bf 75} (2015), no.~7 299,
  [\href{http://arxiv.org/abs/1502.01518}{{\tt arXiv:1502.01518}}]. [Erratum:
  Eur.Phys.J.C 75, 408 (2015)].

\bibitem{Aaboud:2017phn}
{\bf ATLAS} Collaboration, M.~Aaboud et~al., {\it {Search for dark matter and
  other new phenomena in events with an energetic jet and large missing
  transverse momentum using the ATLAS detector}},  {\em JHEP} {\bf 01} (2018)
  126, [\href{http://arxiv.org/abs/1711.03301}{{\tt arXiv:1711.03301}}].

\bibitem{Flacke:2014jwa}
T.~Flacke, K.~Kong, and S.~C. Park, {\it {A Review on Non-Minimal Universal
  Extra Dimensions}},  {\em Mod. Phys. Lett. A} {\bf 30} (2015), no.~05
  1530003, [\href{http://arxiv.org/abs/1408.4024}{{\tt arXiv:1408.4024}}].

\bibitem{Alloul:2013bka}
A.~Alloul, N.~D. Christensen, C.~Degrande, C.~Duhr, and B.~Fuks, {\it
  {FeynRules 2.0 - A complete toolbox for tree-level phenomenology}},  {\em
  Comput. Phys. Commun.} {\bf 185} (2014) 2250--2300,
  [\href{http://arxiv.org/abs/1310.1921}{{\tt arXiv:1310.1921}}].

\end{thebibliography}\endgroup

\end{document}